\documentclass[twocolumn]{aastex63}

\RequirePackage{color}

\newcommand{\ltsima}{$\; \buildrel < \over \sim \;$}
\newcommand{\simlt}{\lower.5ex\hbox{\ltsima}}
\newcommand{\gtsima}{$\; \buildrel > \over \sim \;$}
\newcommand{\simgt}{\lower.5ex\hbox{\gtsima}}
\newcommand{\msun}{{\cal M_\odot}}
\newcommand{\boo}{{Boo~I}}

\definecolor{darkcoral}{rgb}{0.8, 0.36, 0.27}

\graphicspath{{./}{figures/}}

\usepackage{amsmath}

\received{N/A}
\revised{N/A}
\accepted{September 19, 2022}

\shorttitle{Faint Stars in a Faint Galaxy II}
\shortauthors{Filion et al.}

\begin{document}

\title{Faint Stars in a Faint Galaxy: II. The Low Mass Stellar Initial Mass Function of the Bo{\"o}tes I Ultrafaint Dwarf Spheroidal Galaxy}

\correspondingauthor{Carrie Filion}\email{cfilion@jhu.edu}
\author[0000-0001-5522-5029]{Carrie Filion}
\affil{{Department of Physics \& Astronomy, The} Johns Hopkins University, {Baltimore, MD 21218}}

\author{Imants Platais}
\affil{{Department of Physics \& Astronomy, The} Johns Hopkins University, {Baltimore, MD 21218}}

\author[0000-0002-4013-1799]{Rosemary F.G. Wyse}
\affil{{Department of Physics \& Astronomy, The} Johns Hopkins University, {Baltimore, MD 21218}}

\author{Vera Kozhurina-Platais}
\affil{Space Telescope Science Institute, {Baltimore, MD 21218}}





\begin{abstract}
This paper presents improved constraints on the low-mass stellar initial mass function (IMF) of the Bo{\"o}tes I (Boo~I) ultrafaint dwarf galaxy, based on our analysis of recent deep imaging from the Hubble Space Telescope. The identification of candidate stellar members of Boo~I in the photometric catalog produced from these data was achieved using a Bayesian approach, informed by complementary archival imaging data for the Hubble Ultra Deep Field. Additionally, the existence of earlier-epoch data for the fields in Boo~I allowed us to derive proper motions for a subset of the sources and thus identify and remove likely Milky Way stars. We were also able to determine the absolute proper motion of Boo~I, and our result is in agreement with, but completely independent of, the measurement(s) by \textit{Gaia}. The best-fitting parameter values of three different forms of the low-mass IMF were then obtained through forward modeling of the color-magnitude data for likely Boo~I member stars within an approximate Bayesian computation Markov chain Monte Carlo algorithm. The best-fitting single power-law IMF slope is $\alpha = -1.95_{-0.28}^{+0.32}$, while the best-fitting broken power-law slopes are $\alpha_1 = -1.67_{-0.57}^{+0.48}$ and $\alpha_2 = -2.57_{-1.04}^{+0.93}$. The best-fitting lognormal characteristic mass and width parameters are $\rm{M}_{\rm{c}} = 0.17_{-0.11}^{+0.05} \msun$ and $\sigma=0.49_{-0.20}^{+0.13}$. These broken power-law and lognormal IMF parameters for Boo~I are consistent with published results for the stars within the Milky Way and thus it is plausible that Bo{\"o}tes I and the Milky Way are populated by the same stellar IMF.

\end{abstract}



\section{Introduction} \label{sec:intro}

The stellar initial mass function (IMF) describes the mass distribution of single stars formed in any given star formation event. Stars of different masses have different impacts on their environment and play distinct roles in many aspects of galaxy evolution and the cosmic baryon cycle. At the high-mass end, for example, the IMF sets the number of core-collapse supernovae that can occur in a system of given total stellar mass, which determines (e.g.) the level of chemical enrichment and amplitude of possible early stellar feedback. At the low-mass end ($M~\lesssim~1~\msun$), the IMF determines the number of long-lived stars in a system that \lq lock up' baryons on long timescales - which, in the Milky Way, comprise the majority of the total stellar mass. 

There are a variety of approaches to determining both the high and low-mass ends of the IMF discussed in the literature. In this paper we discuss only star-count based analyses of the low-mass IMF. Due to the long main sequence lifetimes of low-mass stars, essentially all the stars in an ancient stellar system that formed below the main sequence turn-off (MSTO) are still alive today. The present day mass function (PDMF, determined through star counts) is thus closely related to the IMF, and the low-mass IMF can in principle be determined from the PDMF after corrections for photometric incompleteness, unresolved binary systems, etc. are properly modelled and applied.

The low-mass IMF of the stellar populations of the Milky Way, as constrained in this way in the bulge, the disk(s), and the local stellar halo (i.e. populations with different age and metallicity distributions), appears to be consistent with being invariant (see e.g. the review of \citealt{bastian_2010}). The early, influential work of \cite{salpeter_1955} describes the local IMF from $0.4~\msun$ to $10~\msun$ as a single power law with slope $\alpha = -2.35$. The Salpeter single power-law slope has been used, over all masses, to describe both single-star and system IMFs, where the system IMF is that inferred from star counts when binary star systems are treated as single sources. However, studies in the decades since with improved data sets have found that the low-mass IMF shows a turn-over at the low-mass end, below $\sim 1 \msun$. This (invariant) IMF is well-described by either a broken power law or a lognormal function. For example, the \cite{kroupa_2001} single-star broken power-law IMF has a break at $\rm{M}_{\rm{BK}}~=~0.5~\msun$ and slopes of $\alpha_{\rm{1K}} = -1.3$ and $\alpha_{\rm{2K}} = -2.3$ below and above the break, respectively, and the \cite{chabrier_2005} lognormal IMF\footnote{The often-quoted values of $\rm{M}_{cc} = 0.22 \msun$, $\sigma_{\rm{C}_{\rm{M}}} = 0.57$ are for the \textit{system} IMF presented in \cite{chabrier_2003}} has a characteristic mass of $\rm{M}_{\rm{CC}}~=~0.2~\msun$ and width parameter $\sigma_{\rm{C}_{\rm{M}}}~=~0.55$.

Star-count based analyses of the IMF in nearby ultrafaint dwarf (UFD) galaxies, typically defined as dark-matter dominated systems with $\rm{L} \lesssim 10^5 \rm{L}_\odot$ (see e.g. \citealt{simon_review} for review), however, have provided evidence that the IMF(s) in these extremely low stellar mass, low-metallicity galaxies may vary. The first such study, by \cite{geha_2013}, used photometry from Hubble Space Telescope (HST) Advanced Camera for Surveys Wide Field Camera (ACS/WFC) imaging to constrain the IMF of the Hercules (Herc) and Leo IV UFD galaxies, over the mass range $\sim~0.5\msun$ to $\sim~0.8~\msun$. Those authors fit both single power law and lognormal single-star IMFs to their data, and find that their best-fit single power-law slope values are shallower than the Salpeter IMF value (and the Kroupa value above the break mass), while their lognormal characteristic mass contains the Chabrier value within the one-sigma confidence interval (holding the lognormal $\sigma$ fixed at $0.69$). Combined with literature results from higher luminosity systems (including the Milky Way), the authors propose a possible trend of increasingly bottom-light IMFs with decreasing metallicity and/or velocity dispersion. Further, \cite{gennaro_2018a} use ACS/WFC photometry to constrain the \textit{system} IMFs of the Bo{\"o}tes I (\boo), Canes Venatici II, Coma Berenices (ComBer), Herc, Leo IV, and Usra Major I UFD galaxies, over the mass range $\sim~0.45~\msun$ to $\sim~0.8~\msun$. They find that the best-fitting single power-law slopes and lognormal IMF parameters vary among the galaxies in the sample, and again that the best-fitting single power-law slopes are shallower than the Salpeter slope. The values of the \textit{system} lognormal characteristic mass that they determine for each of the galaxies in their sample are higher than that of the Chabrier \textit{system} IMF, but the values of the width parameter ($\sigma$) that they determine are similar to that of the Chabrier \textit{system} IMF. These authors also find a possible trend of shallower IMF slope with decreasing metallicity, and a weaker possible trend with velocity dispersion. We note that both \cite{geha_2013} and \cite{gennaro_2018a} establish that the mass distributions of UFD galaxies do not need to be corrected for internal dynamical effects, as the relaxation time in these dark-matter dominated systems is longer than the age of the Universe.

However, as discussed in \cite{elbadry_2017}, fitting a single power-law IMF, constrained only over the narrow mass range $\sim~0.5~\msun$ to $\sim~0.8~\msun$ (such as those analyses discussed above), to a system with a true underlying IMF that is lognormal can produce an artificially shallow slope. These authors also show that when constraining the lognormal form, observations must reach to approximately the characteristic mass to be able to strongly constrain both the characteristic mass and width parameter (which, in the stellar populations of the Milky Way, would require that observations reach $\sim 0.2\msun$, \citealt{chabrier_2005}). As such, it is paramount to perform IMF analyses with observational data that reach to lower masses ($M < 0.5 \msun$), especially if the ultimate goal is comparison to the IMF of the stellar populations of the Milky Way.

The deepest analysis of the IMF in a UFD galaxy to date is that presented in \cite{gennaro_2018b}, who revisit the determination of the low-mass (system) IMF of ComBer, this time using infrared photometry from the HST Wide Field Camera~3, probing masses down to $\sim 0.2 \msun$. Their results are consistent with those for ComBer based on shallower optical data given in \cite{gennaro_2018a}, and the broken power law and lognormal parameters that they determine are consistent with the Kroupa (single-star) and Chabrier (system) IMF, respectively. Their best-fit single power-law slope is inconsistent with the Salpeter slope, however. This inconsistency is perhaps unsurprising, as over the mass range probed in their analysis, the Salpeter IMF is an inappropriate parameterization for even the IMF of the stellar populations of the Milky Way. Their analysis underscores the importance of using sufficiently deep data for the determination of the low-mass IMF, and supports the possibility that (at least some) UFD galaxies may have the same IMF as that of the stellar populations in the Milky Way, despite their otherwise extremely different properties (such as estimated total mass).

In this paper, we present a new investigation of the low-mass single-star IMF of \boo, utilising the ultra-deep photometry presented in \citet[hereafter Paper~I]{paper1}. The limiting magnitude of these data correspond to a stellar mass of $\sim 0.3 \msun$, approximately $\sim 0.15 \msun$ less massive than the limit reached by the \boo\ data analyzed in \cite{gennaro_2018a}. These new deeper data enable an independent analysis of the low-mass IMF of \boo\ that probes a lower mass regime. We describe the photometric data employed in this work in Section~\ref{sec:data}, and outline how we created the catalog of likely member stars of \boo\ using these data in Section~\ref{sec:membership}. We then explain our methodology for constraining the low-mass IMF from the catalog of likely member stars and present results in Section~\ref{sec:IMF}. We discuss these results in the context of the present literature in Section~\ref{sec:discussion}, and we conclude in Section~\ref{sec:conclusion}.


\section{Observational Data} \label{sec:data}

In this analysis we used the catalog of DAOPHOT PSF photometry \citep{stetson_1987} obtained in Paper~I. These data were produced from imaging taken under our HST GO program 15317 (PI I. Platais), and the observing strategy, data reduction, and photometric analysis of these data are detailed in Paper~I. This photometric catalog was produced from  HST ACS/WFC imaging of three slightly overlapping fields (or pointings) centered on \boo, taken using the F606W and F814W filters. The completeness of these data was determined using artificial star tests performed independently in each filter, as described in Paper~I, which gave a $50\%$ completeness limit of 27.4 and 28.2 for the F814W and F606W filters, respectively, in the Vega magnitude system.

For the purposes of this work, we made two further enhancements to the catalog presented in Paper~I. The release of \textit{Gaia} EDR3 (\citealt{gaia_2016}, \citealt{gaiaedr3}) allowed us to calibrate the celestial coordinates to this frame, as detailed in Appendix \ref{sec:astrm}. This calibration made it easier to identify sources that had multiple entries, such as those on the overlap regions of two different pointings, in the original, DAOPHOT PSF photometric catalog. We averaged the photometry for duplicated sources, and ensured only unique sources were kept in the DAOPHOT photometric catalog.

The issue of contamination by non-stellar sources (e.g. galaxies) was addressed in Paper~I by applying cleaning cuts based on the statistics of the artificial star tests. These cuts resulted in a \lq cleaned' photometric catalog that contained $\sim 3,000$ sources. In the present analysis, we derived an alternative Bayesian approach that allowed us to simultaneously remove non-stellar sources and identify likely stellar members of \boo\ in the uncleaned photometric catalog (we refer to this catalog, which contains $\sim 8,000$ sources, as \lq the \boo\ field data' or the \lq the \boo\ field catalog'). This new Bayesian approach required the characterisation of the statistical properties of galaxies and other non-stellar sources, which we achieved through an analysis of the Hubble Ultra Deep Field (HUDF). The number of expected Milky Way foreground contaminants in each of the HUDF and the \boo\ field is expected to be moderate due to their high Galactic latitudes, and the HUDF should be dominated by non-stellar sources. We produced photometry for a single ACS/WFC pointing in the HUDF following an identical procedure to that presented in Paper~I, and we turn to this analysis next.

\subsection{The Hubble Ultra Deep Field Photometric Catalog}\label{sec:udf_phot}

We chose exposures from GO-11563 (PI G. Illingworth), as their observing strategy allowed us to create a catalog of similar depth to the \boo\ field data. We selected individual ACS/WFC F814W and F606W images that had exposure times that were similar to those of the our \boo\ field imaging, and further required that the total exposure times in each filter sum to approximately that of each pointing in \boo\ field. This resulted in eight F814W exposures ($4~\times~1307$~s and $4~\times~1225$~s) and seven F606W exposures ($3~\times~1469$~s and $4~\times~1331$~s). We did not perform a separate analysis of the completeness in this field, as it should be comparable to the \boo\ field.

We reduced these data following the procedure for DAOPHOT PSF photometry given in Paper~I, adopting the identical parameter values and the analytic PSF model fit that was determined from the \boo\ field data. 
The HUDF has lower line-of-sight extinction than the \boo\ field, and we thus added extinction to the HUDF photometry to match the \boo\ field. We determined the appropriate correction using $\rm{E(B-V)} = 0.04$ for the \boo\ field \citep{brown_sfh}, $\rm{E(B-V)} = 0.0205$ for the HUDF (based on the dust maps from \citealt{schlafly_2011}), an $R = 3.1$ extinction curve, and the ratio of extinction in our passbands to the extinction in the V band\footnote{provided by \url{http://svo2.cab.inta-csic.es/theory/fps/}}. 

\begin{figure*}
\includegraphics[width=.95\textwidth]{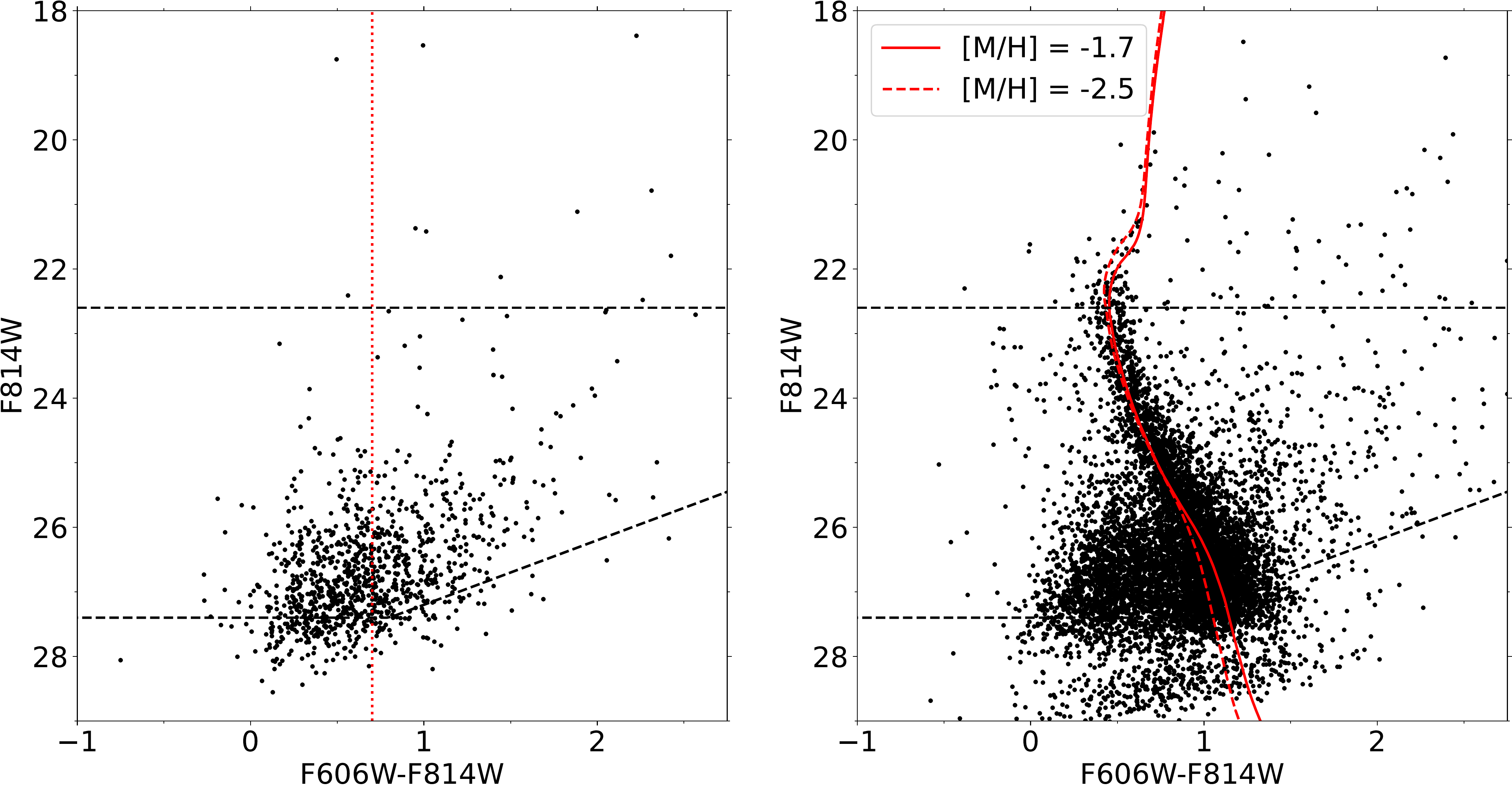}
\caption{CMDS of all sources that were fit with DAOPHOT photometry. Left: the sources in the HUDF PSF catalog. Right: the sources in the \boo\ field, where no quality cuts have been applied. The mean color of the sources in the HUDF PSF catalog is indicated by the vertical dotted line. Two solar-scaled 13~Gyr isochrones ($[\rm{Fe}/\rm{H}] = [\rm{M}/\rm{H}] = -1.7$, solid, and $[\rm{Fe}/\rm{H}] = [\rm{M}/\rm{H}] = - 2.5$, dashed) from the Dartmouth Stellar Evolution Database \citep{dartmouth} are overplotted on the \boo\ data, after adjustment for distance and extinction. The two metallicities shown correspond to the approximate mean and maximum metallicity in \boo, and while these isochrones overlap at brighter magnitudes, they diverge at the faint (low mass) end, illustrating the importance of metallicity information. In both panels, the dashed black lines show our imposed magnitude limits for the final \boo\ catalog used to constrain the IMF.}

\label{fig:udf_cmd}
\end{figure*}

We refer to the resulting photometric catalog ($\sim 1000$ sources in total) as the \lq HUDF PSF catalog', and assume that it is composed almost entirely of galaxies and other non-stellar sources. This assumption of low stellar contamination is supported by the analysis presented in \cite{pirzkal_2005}, which identified only $\sim~50$ total point sources in the HUDF and confirmed that $\sim 30$ of these sources (with AB magnitude $\rm{F775W} \le 27$) were stars or white dwarf stellar remnants. It is important to note that this \lq HUDF PSF catalog' likely has different properties than other HUDF catalogs in the literature, as this catalog is only intended to be used for the purposes stated above, rather than to facilitate study of galaxy populations. The CMDs of the \boo\ field catalog and the HUDF PSF catalog are presented in Figure~\ref{fig:udf_cmd}.


\section{Identification of Likely Member Stars of Bo{\"o}tes I} \label{sec:membership}

\subsection{Photometric Membership Determination} \label{sec:bayesian_cleaning}

From inspection of Figure~\ref{fig:udf_cmd}, it is evident that many of the stars in the \boo\ field follow an ancient, metal-poor isochrone, unlike the HUDF PSF sources. Indeed, semi-resolved background galaxies are known to be, on average, bluer than stars on the old, MSTO of \boo\ (see e.g. \citealt{tyson_1988}, \citealt{bedin_2008}). Further, galaxies should not be well fit by a PSF model, and thus they can be distinguished from stars via statistics related to the quality of the PSF fit (in this case, the DAOPHOT \textit{sharp} statistic, which is measure of the difference between the width of the PSF and the width of the source). However, stars and galaxies have overlapping distributions of both color and \textit{sharp} statistics, especially towards fainter magnitudes, and even stars can occasionally be poorly fit by the PSF, which complicates the identification and removal of likely non-stellar sources. Cleaning cuts based on adopted threshold values of some combination of statistics and/or color information are usually employed to remove non-stellar objects, as was done in Paper~I. However, as discussed in Paper~I, faint galaxies can still remain after the application of stringent statistics-based cuts. Milky Way stars will also remain, although their color at a given magnitude will usually be inconsistent with \boo\ membership, and thus additional color-magnitude based cuts are typically required to remove Milky Way stellar contaminants.

In the present analysis we developed an alternative technique. We employed a Bayesian approach that probabilistically identified the sources that have both photometric quality-of-fit statistics and colors at a given magnitude consistent with their being stellar members of \boo. This approach allowed us to simultaneously identify and remove both non-stellar sources and likely Milky Way stars, thus avoiding the traditional application of a series of harsh, threshold-based cuts. We refer to the sources that are not likely stellar members of \boo\ as \lq non-member sources' (NMSs). These NMSs are a combination of galaxies and Milky Way stars, and we note that all $\sim 1000$ sources in the HUDF PSF catalog are NMSs.

According to Bayes' theorem, the probability that a source is a member star or NMS, given some data for the source (the posterior probability), can be written as: 

\begin{equation}
P(\rm{T}_{i} | \rm{data}) = \frac{P(\rm{data} |\rm{T}_{i}) P (\rm{T}_{i})} {P(\rm{data})},
\end{equation}
where $\rm{T}_{i} = \rm{T}_1$ for a member star and $\rm{T}_{i} = \rm{T}_2$ for a NMS. The data that we considered for each source consisted of its $\rm{F606W} - \rm{F814W}$ color {($\rm{C}_{\rm{M}}$, where here the subscript \lq M' denotes that color is a function of magnitude)}, its F814W DAOPHOT \textit{sharp} statistic ($\rm{S}_{\rm{F814W}}$), and its F606W DAOPHOT \textit{sharp} statistic ($\rm{S}_{\rm{F606W}}$). As such, the posterior probability became:
\begin{equation}\label{eq:bayes}
\begin{split}
P(\rm{T}_i | \rm{C}_{\rm{M}}, \rm{S}_{\rm{F814W}}, \rm{S}_{\rm{F606W}}) = \\
\frac{P(\rm{C}_{\rm{M}}, \rm{S}_{\rm{F814W}}, \rm{S}_{\rm{F606W}} | \rm{T}_i)P(\rm{T}_i)} {P(\rm{C}_{\rm{M}}, \rm{S}_{\rm{F814W}}, \rm{S}_{\rm{F606W}})}.
\end{split}
\end{equation}
We assumed that $\rm{C}_{\rm{M}}, \rm{S}_{\rm{F814W}}$, and $\rm{S}_{\rm{F606W}}$ are {conditionally} independent  {(i.e. given that a source is a star, knowledge of $\rm{S}_{\rm{F814W}}$ provides no information about $\rm{S}_{\rm{F606W}}$ or $\rm{C}_{\rm{M}}$, for example)}, and thus the right hand side of Equation~\ref{eq:bayes} could be expanded into:

\begin{equation}\label{eq:full_bayesian}
\frac{P(\rm{C}_{\rm{M}}| \rm{T}_i) P(\rm{S}_{\rm{F814W}} | \rm{T}_i) P(\rm{S}_{\rm{F606W}} | \rm{T}_i) P(\rm{T}_i)}{\sum\limits_{i} P(\rm{C}_{\rm{M}}| \rm{T}_i) P(\rm{S}_{\rm{F814W}} | \rm{T}_i) P(\rm{S}_{\rm{F606W}} | \rm{T}_i) P(\rm{T}_i)}.
\end{equation}

Note that the denominator ensured that the total probability was normalized to unity (i.e. $P(\rm{T}_1 | \rm{C}_{\rm{M}}, \rm{S}_{\rm{F814W}}, \rm{S}_{\rm{F606W}})$ + $P(\rm{T}_2 | \rm{C}_{\rm{M}}, \rm{S}_{\rm{F814W}}, \rm{S}_{\rm{F606W}}) = 1)$. We then determined $P(\rm{C}_{\rm{M}}| \rm{T}_i)$, $P(\rm{S}_{\rm{F814W}} | \rm{T}_i)$, $P(\rm{S}_{\rm{F606W}} | \rm{T}_i)$, and $P(\rm{T}_i)$, as discussed below.

The member stars of \boo\ should have star-like \textit{sharp} statistics, and a color-magnitude distribution akin to an isochrone representative of an ancient, metal-poor population. Here, we adopt a fiducial isochrone from the Dartmouth Stellar Evolution Database \citep{dartmouth} with properties representative of \boo, i.e. age of 13~Gyr, $[\rm{Fe}/{H}] = -2.5$, {{appropriately}} adjusted for a distance\footnote{This is more distant than was assumed in Paper~I, but is a good fit to the blue horizontal branch population \citep{filion_wyse} and is consistent with RR Lyrae-based distance estimates \citep{siegel_2006, dallora_2006}} of $65$~kpc \citep{okamoto_2012} and extinction (see Section~\ref{sec:udf_phot}). We used the data from the artificial star tests (described in Paper~I) and this fiducial isochrone to determine the probability distributions {of color at a given observed apparent magnitude, $\rm{S}_{\rm{F814W}}$, and $\rm{S}_{\rm{F606W}}$ for  stellar members of \boo}. The NMSs in the \boo\ field catalog should be the same types of sources as those in the HUDF PSF catalog, and thus we used the HUDF PSF catalog to determine the probability distributions of {color at a given apparent magnitude, $\rm{S}_{\rm{F814W}}$, and $\rm{S}_{\rm{F606W}}$ for NMSs.}

We first determined the probability distributions of the \textit{sharp} parameters  of stellar members and NMSs by fitting Gaussians to the (non-binned) $\rm{S}_{\rm{F606W}}$ and $\rm{S}_{\rm{F814W}}$ data from the artificial star test catalogs and the HUDF PSF catalog, respectively. We then assumed that the color of {a member star of \boo\ is a function of its} $\rm{F814W}$ apparent magnitude, as given by the fiducial isochrone described above, {and used this isochrone to determine the mean of the Gaussian probability distribution of the color at that magnitude}. We interpolated along this isochrone to the $\rm{F814W}$ apparent magnitude of each source in the \boo\ field catalog, and adopted the $\rm{F606W} - \rm{F814W}$ color of the isochrone at that apparent magnitude as the mean of the Gaussian probability distribution for {color}. {We adopted a constant value for the Gaussian standard deviation, which was informed by the mean (over all magnitudes) of the standard error in each measurement of magnitude from the artificial star tests, again remembering that these tests were performed independently in each filter. Specifically, we determined the means of the standard errors in $\rm{F814W}$ and $\rm{F606W}$, added these means in quadrature and then  multiplied by a factor of three to obtain the standard deviation, which equaled   $0.09$~mag. Over the range of $\rm{F814W}$ apparent magnitudes analyzed, the equal-mass binary sequence is separated from the single-star isochrone by a $\rm{F606W} - \rm{F814W}$ color of between $\sim 0.05$ and $\sim 0.14$, and thus this adopted standard deviation value is sufficiently large that unresolved binary systems are incorporated within our probabilistic framework.}

\begin{figure*}
\includegraphics[width=1\textwidth]{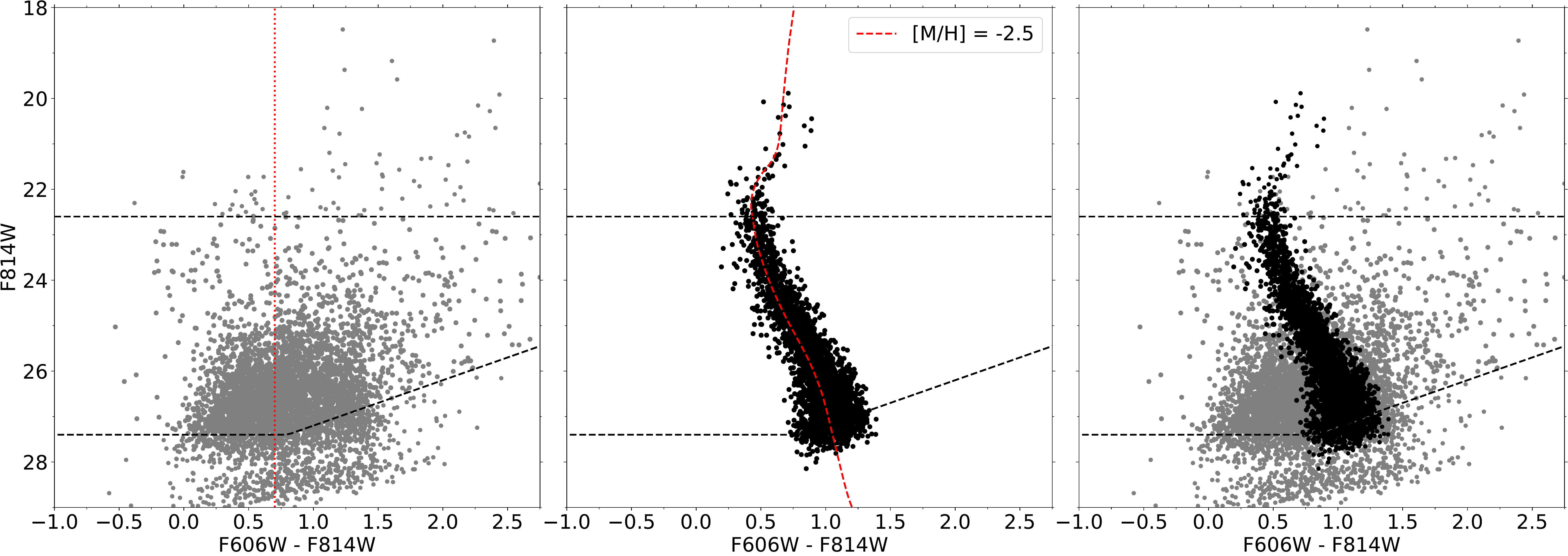}
\caption{CMDs of the non-member sources and likely member stars in the \boo\ field, as determined by the Bayesian probabilities computed in Section~\ref{sec:bayesian_cleaning} and the adopted $50\%$ probability threshold. Left: all non-member sources in the \boo\ field. Center: {all candidate \boo\ member stars. The isochrone is as described in Figure~\ref{fig:udf_cmd}}. Right: all sources (non-member sources shown in grey, candidate member stars in black). {In all panels, the dashed black lines show our imposed magnitude limits for the final \boo\ catalog used to constrain the IMF.}}
\label{fig:star_gal_cmd}
\end{figure*}

As seen in Figure~\ref{fig:udf_cmd}, the colors of sources in the HUDF PSF catalog do not strongly depend on magnitude (note that the roughly diagonal trend at redder color, fainter than $F814W \sim 27$, is due to photometric incompleteness). We thus modeled the probability distribution of the colors of NMSs as a Gaussian that is independent of apparent magnitude, with the mean ($0.70$) and standard deviation ($0.44$) determined via a fit to the (non-binned) color distribution. This mean color value is shown as a vertical line in Figure~\ref{fig:udf_cmd} and \ref{fig:star_gal_cmd}.

Finally, we determined the priors, $P(\rm{T}_1)$ and $P(\rm{T}_2)$, from the ratio of the number of sources in the HUDF PSF {to that in the \boo\ field catalog, taking into account} that the \boo\ field has three times the areal coverage. {Here, the priors reflect the overall relative populations of stars and NMSs (i.e. the probabilities  of being a star or NMS, without any additional information). \boo\ and the HUDF are both at relatively high Galactic latitudes\footnote{The Galactic coordinates of the HUDF are $(\ell, b) \simeq (223^\circ, -54^\circ)$ and those of \boo\ are $(\ell, b) \simeq (358^\circ, 70^\circ)$}, albeit at somewhat different longitudes.  The longitude of \boo\ is more towards the Galactic center while  the HUDF line-of-sight lies closer to the Galactic plane  and it may be expected that these fields have similar numbers of Milky Way stars. This expectation is supported by the similarity between the number of point sources identified in the HUDF analysis presented \citealt{pirzkal_2005} (totalling $\sim 50$), and the number of predicted Milky Way sources in the \boo\ field ($\sim 140$ in an area that is three times larger than the HUDF), see Section~\ref{sec:pm_identification} below, and we proceed under the assumption that the number of Milky Way stars is similar in each field. Further, at these high Galactic latitudes, the NMSs will be mostly galaxies, which should have approximately constant density on the sky.} We then estimated $P(\rm{T}_2)~=~\sim~\frac{3 \times 1000}{8000} \sim 40\%$, and $P(\rm{T}_1) = 1 - P(\rm{T}_2) \sim 60\%$. We note that these priors are approximate, and we verified that altering these prior values has minimal effect on the final number of probable member stars. For example, increasing $P(\rm{T}_2)$ by fifty percent (to $60\%$) changes the number of likely member stars (defined as $P(\rm{T}_1 | \rm{C}_{\rm{M}}, \rm{S}_{\rm{F814W}}, \rm{S}_{\rm{F606W}}) > 50\%$) within the magnitude limits given below by less than $5\%$.

We then followed Equation \ref{eq:full_bayesian} and computed $P(\rm{T}_1 | \rm{C}_{\rm{M}}, \rm{S}_{\rm{F814W}}, \rm{S}_{\rm{F606W}})$ and $P(\rm{T}_2 | \rm{C}_{\rm{M}}, \rm{S}_{\rm{F814W}}, \rm{S}_{\rm{F606W}})$ for each source in the \boo\ field using these probability distributions and priors. After experimentation, we adopted a $50\%$ probability threshold and we considered a source to be a likely member star if $P(\rm{T}_1 | \rm{C}_{\rm{M}}, \rm{S}_{\rm{F814W}}, \rm{S}_{\rm{F606W}})~>~50\%$, or a NMS if  $P(\rm{T}_2 | \rm{C}_{\rm{M}}, \rm{S}_{\rm{F814W}}, \rm{S}_{\rm{F606W}})~>~50\%$. We present CMDs of the resulting classifications of sources in the \boo\ field in Figure~\ref{fig:star_gal_cmd}. The leftmost CMD shows all NMSs, the middle CMD shows all likely member stars, and the rightmost CMD shows both likely member stars and NMSs. The adopted bright and faint limits for the \boo\ photometric catalog ($22.6 \le F814W \le 27.4$, $F606W \le 28.2$) are shown in all panels, reaching from slightly fainter than the apparent MSTO to the $50\%$ completeness limits of the photometry. {Of the initial $\sim 8,000$ total sources, $\sim 3,000$ were identified as likely member stars of Boo~I via the Bayesian methodology adopted here, and $2,586$ of these likely member stars were within the adopted apparent magnitude limits to be considered in the IMF determination} As discussed in Paper~1, using a symmetric color distribution about a single isochrone to determine likely photometric member stars has the potential to misidentify unresolved binary systems as non-members. However, visual inspection of the CMDs in Figure~\ref{fig:star_gal_cmd} indicates this not to be the case, reflecting the fact that the Bayesian membership probabilities use a relatively wide color distribution and incorporate non-color based factors.


The number of likely member stars found via this new  approach is approximately the same as that obtained in Paper~I, but the cloud of faint blue, likely galaxy sources seen in the lower left of the CMDs of Paper~I has now been removed. The candidate member stars have colors and apparent magnitudes consistent with those expected for stellar members of \boo, as anticipated. However, it is possible for Milky Way stars to fall within this locus on the CMD; the majority of these stars should be members of the stellar halo, and hence are likely to be distinguishable through proper motions, as discussed in Section~\ref{sec:pm_identification}.

\subsubsection{Estimation of Contamination in the Candidate Member Catalog}\label{sec:contamination}

First we made an empirical estimate by computing the probability of being a member star given the data ($P(\rm{T}_1|\rm{data})$) for each of the sources in the HUDF PSF catalog, from which we estimated the number of non-member sources that could be included in the candidate member catalog. We identified 25 sources in the HUDF PSF catalog with probability of membership greater than $50\%$ (i.e. $P(\rm{T}_1|\rm{C}_{\rm{M}}, \rm{S}_{F814W}, \rm{S}_{F606W}) > 50\%$); 13 of which are within our magnitude limits, and all 13 are at the lower signal-to-noise faint end ($F814W \gtrsim 24.2$). This low number indicates that $\frac{25}{\sim 1000} \lesssim 3\%$ of the HUDF PSF catalog sources (NMSs) make it through the selection of candidate members. The catalog of probable \boo\ members should contain a similarly low level of non-member contamination, predominantly consisting of galaxies and a few stars.

We then used the analytic TRILEGAL Milky Way models \citep{trilegal} to estimate the likely number of Milky Way stars in the \boo\ field that have colors and apparent magnitudes that place them near the fiducial isochrone. The model predicts a total of $\sim 140$ total Milky Way stars in the \boo\ footprint, and that $\sim 60$ of these should have (error-free, extinction-free) magnitudes within our adopted magnitude limits. Of these, $\sim 20$ stars have $\rm{F606W}~-~\rm{F814W}$ colors within 0.5 mag of the fiducial isochrone, placing them close to the locus occupied by likely members of \boo. As discussed in Appendix \ref{sec:astrm}, we were able to measure proper motions for a subset of the sources in the \boo\ field. We can then anticipate that some fraction of the predicted $\sim 20$ Milky Way stars can be identified via their proper motions, and we turn now to the astrometric identification of non-member stars.

\subsection{Absolute Proper Motion of Bo{\"o}tes~I}\label{sec:pm_boo}

We matched the field centers from GO-12549 (PI: T. Brown) so that proper motions could be attempted using a baseline of $\sim 7$ years. A detailed description of the derivation of the proper motions from these two epochs and the creation of a catalog of astrometric sources in given in Appendix \ref{sec:astrm}. From these proper motion data, we determined the absolute proper motion of \boo: $\mu_{\alpha \ast}$=0.42$\pm$0.04~mas~yr$^{-1}$ and $\mu_{\delta}$=1.00$\pm$0.04~mas~yr$^{-1}$. This measurement is entirely consistent with the results from Gaia (e.g. \citealt{filion_wyse}, \citealt{pace_2022} and references therein). Each of these two measurements use different techniques and datasets and their agreement was not guaranteed, so this consistency is extremely exciting.

\subsection{Proper Motion Identification of Non-Member Stars}\label{sec:pm_identification}

Here, we describe how we used these proper motions to remove likely Milky Way stars from the catalog of candidate \boo\ members found above. It is important to note that the  DAOPHOT PSF photometric catalog extends to fainter magnitudes than the astrometric catalog, and it only contains sources for which a PSF fit was possible in both filters. As such, each catalog contains sources that the other does not\footnote{The fraction of astrometric sources lacking a counterpart in the DAOPHOT photometric catalog increases towards fainter magnitudes. Given that all sources in the DAOPHOT photometric catalog must be fit by a PSF in both filters, the sources that are missing from this catalog are likely semi-resolved galaxies.}.

\begin{figure*}
\includegraphics[width=1\textwidth]{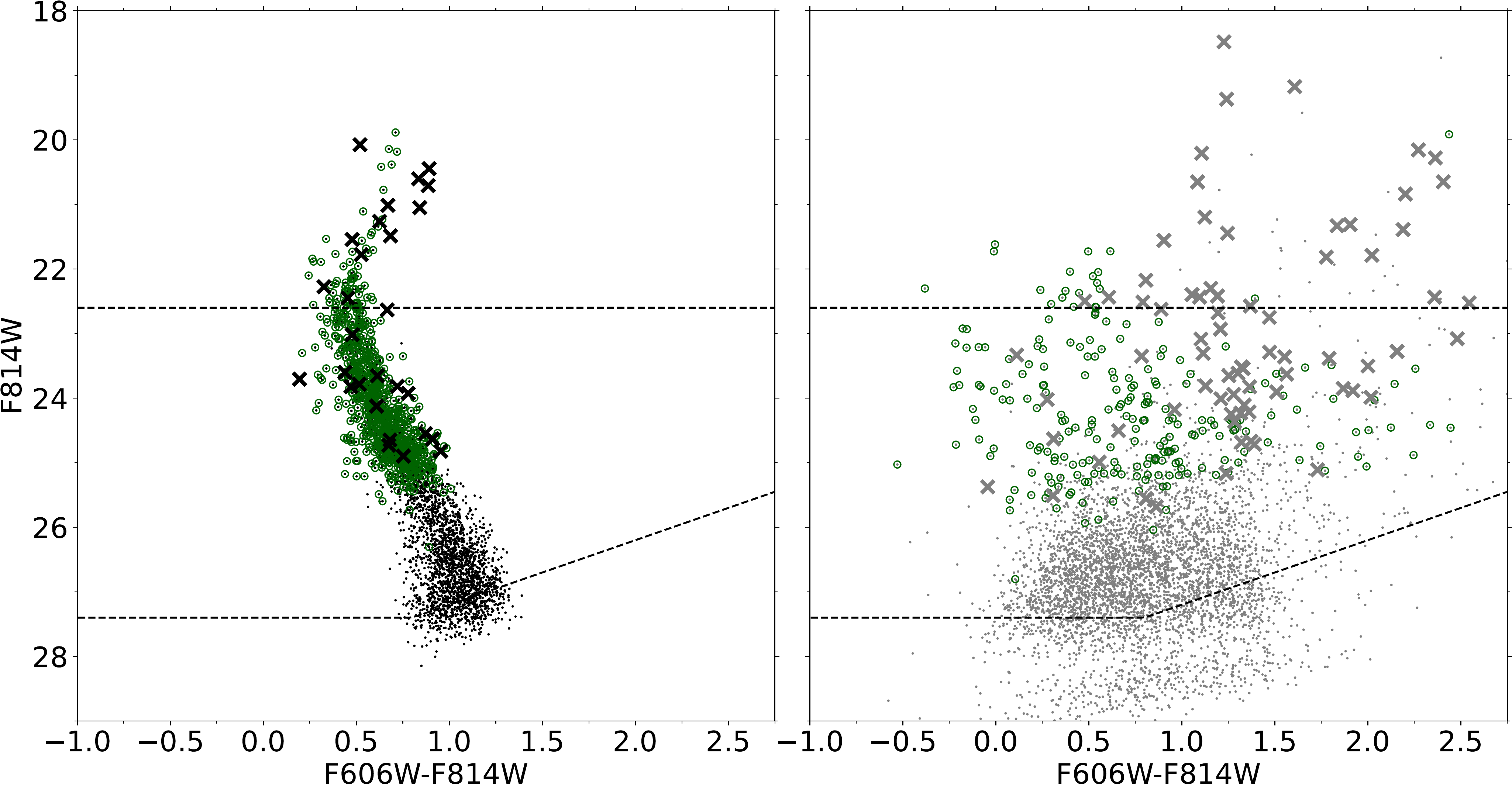}
\caption{CMDs of likely member stars and non-member sources (as determined with the Bayesian probabilities computed in Section~\ref{sec:bayesian_cleaning}) alongside the subset of sources with both this photometry-based Bayesian membership determination and astrometric information. Left: likely member stars. Right: likely non-member sources. In both panels, crosses show sources with DAOPHOT photometry that have proper motions inconsistent with that of \boo, and open green circles indicate sources for which the measured proper motion is consistent with \boo\ (within generous $5\sigma$ error bounds). Sources without proper motion measurements are shown as small points. Dashed lines show magnitude limits of the final catalog used to constrain the IMF.}
\label{fig:star_gal_cmd_pm}
\end{figure*}

We first considered only the sources in the DAOPHOT photometric catalog that had the most reliable proper motions in the astrometric catalog (i.e. $M_{astrm}~<~25$, see Figure~\ref{fig:vpd}).
{We then identified obvious non-members of \boo\ based on significantly discrepant proper motions. We adopted generous five-sigma error bounds on the individual measurements of proper motion to maximize the retention of member stars, and we defined a source to have ‘significantly discrepant’ proper motion if the mean proper motion of \boo\ was not contained within the proper motion measurement of the source plus or minus five times its error ($\mu_{\alpha*} \pm 5 \mu_{err, \alpha*}$, $\mu_{\delta} \pm 5 \mu_{err, \delta}$)}. A total of 109 such proper-motion inconsistent sources were identified in this way, 28 of which had previously been found to have photometric properties consistent with membership, from the Bayesian analysis of Section~\ref{sec:bayesian_cleaning}. These 28 stars are most likely Milky Way foreground stars, and are indicated by a large, black X on Figure~\ref{fig:star_gal_cmd_pm}. Sixteen of these 28 non-member stars are within our adopted magnitude limits - in excellent agreement with the predictions of TRILEGAL given above ($\sim 20$ stars). The remaining 81 non-member sources are shown as grey X's in the right-hand panel.


We then created the final catalog of likely member stars of \boo, hereafter ``the \boo\ member catalog". All stars in this catalog have Bayesian probabilities consistent with membership ({i.e.} $P(\rm{T}_1|\rm{data})~>~50\%$), and have apparent magnitudes within our adopted limits ($22.6~\le~F814W~\le~27.4$, and $F606W~\le~28.2$). All stars with measured proper motions inconsistent with \boo\ were removed. This final catalog contains 2570 stars (970 stars with no proper motion measurements, and 1600 with proper motions measurements - albeit some with large errors). We note that this catalog contains both single stars and binary systems, and to highlight this fact, we occasionally refer to the contents of this catalog as \lq sources'. This catalog provides the input to the determination of the low-mass IMF. The full photometric catalog, including the computed $P(\rm{T}_1|\rm{data})$ and $P(\rm{T}_2|\rm{data})$ values, made publicly available online, and we present an outline of the catalog contents in Table \ref{tab:phot_cat}.

\begin{deluxetable*}{llcl}
\tabletypesize{\footnotesize}
\tablecolumns{4}
\tablewidth{0pt}
\tablecaption{Contents of Photometric Catalog \label{tab:phot_cat}}
\tablehead{\colhead{Column} & \colhead{Unit}  & \colhead{Label} & \colhead{Explanations}}
\startdata
0 & ---    &   ID     & Number, {photometric catalog ID} \\
1 & mag &   MAG$\_$F606W    & DAOPHOT PSF photometric magnitude in the F606W filter \\
2 & mag &   MAG$\_$F814W    & DAOPHOT PSF photometric magnitude in the F814W filter \\
3 & mag & err$\_$F606W     & error of the F606W magnitude \\
4 & mag & err$\_$F814W    & error of the F814W magnitude \\
5 & --- & sharp$\_$F606W     & \textit{sharp} statistic in the F606W filter \\
6 & --- & sharp$\_$F814W    & \textit{sharp} statistic in the F814W filter \\
7 & deg    &   RA  & right ascension, decimal degrees (J2000)\\
8 & deg    &   DEC  & declination, decimal degrees (J2000)\\
9 & ---    &   P$\_$mem & Bayesian probability of membership from photometry \\
10 & ---    &  P$\_$nms  & Bayesian probability of non-membership from photometry \\
11 & --- & flag & flag indicating if source meets all criteria for membership \\
& & & ($1 =$ yes, $0 =$ no)\tablenotemark{a}\\
\enddata
\tablenotetext{a}{See last paragraph of Section~\ref{sec:bayesian_cleaning} for criteria} 
\tablecomments{Table is available in machine-readable format at MAST.}
\end{deluxetable*}

\section{Determination of the Low-Mass Initial Mass Function} \label{sec:IMF}
The traditional method of determining the low-mass IMF from photometric data is to first create the present-day luminosity function (LF) for stars below the (oldest) MSTO. After correction for photometric incompleteness, this LF is then converted to the PDMF using stellar models or some suitable mass-luminosity relationship. The PDMF is then transformed into the IMF after accounting for systematics such as unresolved binary systems. However, this method is fraught with uncertainties. For example, photometric errors and unknown individual stellar metallicities make it such that observed magnitudes cannot easily be converted to the true stellar masses. We thus adopted the more robust method of forward modelling, which circumvents these issues (see for example \citealt{geha_2013}, \citealt{gennaro_2018a, gennaro_2018b}, \citealt{sollima_2019}, among others).

In this approach, synthetic data are generated with given input parameter values and are converted into the observational plane. These synthetic data are then compared to the actual observed data. Both the implementation of forward modelling and the choice of metric for comparing the synthetic and observed datasets are non-trivial, and they differ amongst the analyses in the literature. We adopted an Approximate Bayesian Computation (ABC) Markov Chain Monte Carlo (MCMC) algorithm; we describe the generation of synthetic populations in Section~\ref{sec:modelling}, and the ABC MCMC algorithm and metric in Section~\ref{sec:imf_fit} below.

\subsection{Generating Synthetic Color Magnitude Diagrams}\label{sec:modelling}

We begin this section with a brief outline of the procedure for generating synthetic CMDs, and we then detail our assumptions and choices for parameter values in the following paragraphs. {We first had to generate a stellar population, from which we then created a CMD. For each stellar population,} we assigned masses to each \lq star' by drawing stellar masses from an assumed IMF. The total number of drawn \lq stars'  {was always set to be} larger than the number of sources on the observed CMD, as we had to account for {effects such as photometric incompleteness and the presence of unresolved binary systems.} We then randomly paired some fraction of those \lq stars' into binary systems. We next drew metallicities from a metallicity distribution {function that was} consistent with {spectroscopic data for Boo~I}, and randomly assigned these values to the single \lq stars' and binary systems, {such that both members of the binary had the same metallicity.} {We then generated synthetic CMDs of each stellar population. We determined} the absolute magnitude corresponding to each mass and metallicity pair by interpolating within a grid of stellar isochrones of fixed age but different metallicities. We assumed that all binary systems are unresolved and added their fluxes. We converted the absolute magnitudes into apparent magnitudes by applying the distance modulus and the effects of extinction. Finally, we incorporated the effects of photometric error and incompleteness.

\subsubsection{Priors and Drawing Stellar Masses}

For each synthetic population, we drew IMF parameter values and binary fractions ($f_f$) from uniform (flat) prior distributions. We included three different forms of the IMF: a single power law, a broken power law, and a lognormal, each parameterized by the variables given in Section~\ref{sec:intro}. Specifically, the slope ($\alpha$) of the single power-law IMF was drawn from a prior with minimum and maximum bounds of $-4.5$ and $-0.05$. The same prior was assumed for each of the slopes of the broken power-law IMF ($\alpha_1$, $\alpha_2$), while the break mass was held fixed at the Kroupa value (i.e. the commonly accepted value for the Milky Way, $\rm{M}_{\rm{b}} = 0.5\msun$). In the case of the lognormal IMF, we drew the characteristic mass ($\rm{M}_c$) from a prior between $0.05$ and $1 \msun$, and the width parameter ($\sigma$) from between $0.25$ and $1$. Finally, for all functional forms of the IMF, we sampled the binary fraction ($f_f$) from a prior with bounds of 0 (no binaries) and 1 (every star is in a binary).

Once the IMF parameters were fixed, we then drew the number of \lq stars' ($N$) required to approximately match the number of sources on the observed CMD using the functionalities provided in the publicly available code called \texttt{IMF}\footnote{\texttt{IMF} codebase: \url{https://github.com/keflavich/imf}. Note that the lognormal form of the IMF within \texttt{IMF} uses the natural logarithm ($\ln$), as opposed to the base-ten logarithm ($\log$) adopted in (e.g.) \cite{chabrier_2005}. For the same $\sigma$ and $\rm{M}_c$ parameters, the ln form of the lognormal IMF is equivalent to the log form if $\sigma$ in the ln form is replaced with $\sigma \ln{(10)}$. We thus multiplied the drawn $\sigma$ term by $\ln{(10)}$ before drawing masses from the IMF.}. The value of $N$ will vary depending on the binary fraction, and as noted earlier, will be larger than the number of sources on the observed CMD. We drew \lq stars' with a lower mass limit of $0.11 \mathcal{M}_\odot$, which ensured stellar models existed across the range of metallicities considered, and an upper-mass limit of $1 \mathcal{M}_\odot$. This upper limit avoids any assumptions about the high-mass end of the IMF, which we cannot constrain with these data. These mass limits are beyond the range accessible to our observations, but it is important to include stars more and less massive than the observational limits when simulating binary populations, which we turn to next.

\subsubsection{Generating Binary Populations}
First, note that we defined the binary fraction ($f_f$) as the ratio of the number of stars in binary systems to the total number of stars, such that the number of unresolved binary pairs on the CMD is $\frac{N f_f}{2}$. This definition differs from previous studies, such as those by \cite{geha_2013} and \cite{gennaro_2018a, gennaro_2018b}, who defined the binary fraction to be $\frac{\rm{B}}{\rm{S}+\rm{B}}$, where B is the number of binary pairs and S is the number of single stars. Our binary fraction will be systematically higher than theirs, even for the same binary population. Denoting their definition of binary fraction as $f_g$, we find $f_g = \frac{f_f}{2 - f_f}$. 

We then randomly selected $Nf_f$ of the \lq stars' drawn from the IMF to be paired into binary systems. Exactly how stars are paired into binaries needs to be treated with care (see discussion in \citealt{kouwenhoven_2009}). We chose to randomly pair stars from the IMF as it ensures that the single-star IMF inferred from the synthetic CMD is the same as that used to create the population, and it is computationally fast. We note that random pairing may not capture the physics of how binary systems are formed in nature and, indeed, the low-mass binary population of the Milky Way does not appear to be consistent with this pairing scheme (see \citealt{duchene_2013} and references therein for review). {We verified that this choice of pairing scheme does not significantly impact our results, and we present our tests of alternate mass ratio distributions in Appendix \ref{sec:mass_ratio}. We thus proceeded with the simplifying assumption of randomly-paired binary masses.}

Stars of different metallicities evolve at different rates, such that at a fixed age, the mass corresponding to the MSTO varies with metallicity. We thus defined a fixed mass that is above the MSTO but below the tip of the red giant branch for all metallicities considered, and adopted this mass ($0.785 \mathcal{M}_\odot$) as the limit below which we determined photometric magnitudes. {This} ensured that stellar models existed for the range of metallicities at our assumed age. After pairing binaries, we discarded masses higher than this limit (along with any binary companions of those masses). The low and high mass limits for which photometric magnitudes are determined ($0.11 \msun$ and $0.785 \msun$) correspond to (apparent) magnitudes well beyond both the faint and bright limits that we imposed on the photometry (which, at the metallicity of our fiducial isochrone, correspond to $0.280~\msun$ and $0.759~\msun$, respectively). This guaranteed that the mass cuts did not artificially remove stars that could be scattered via photometric error into the apparent magnitude range considered in this work.

\subsubsection{Assigning Metallicities}
All synthetic stars, single and binary, must then be assigned metallicity values prior to the generation of synthetic photometry. Previous spectroscopic analyses have established that the stellar population of \boo\ has a mean iron abundance of $[\rm{Fe}/\rm{H}] \sim -2.5$ and a large spread in both $[\rm{Fe}/\rm{H}]$ (e.g. \citealt{norris_2010b}) and $[\alpha/\rm{Fe}]$ (e.g. the compilation in \citealt{frebel_2016} and references therein). We thus opted to sample from the \textit{total} metallicity ($[\rm{M}/\rm{H}]$) distribution of \boo, combining the $[\alpha/\rm{Fe}]$ and $[\rm{Fe}/\rm{H}]$ from the homogeneous sample of \cite{lai_2011} using the relationship presented in \cite{salaris_2005} (and used in Paper~I). This enabled us to incorporate the overall effect of $\alpha$-enhancement without having to expand the dimensionality of our isochrone interpolation to include a range of $\alpha$-enhancements. We modelled the resulting distribution by a truncated Gaussian with mean $[\rm{M}/\rm{H}] = -2.4$ and standard deviation $\sigma = 0.46$, truncated between $-3.4~\le~[\rm{M}/\rm{H}]~\le~-1.7$. The single \lq stars' and binary pairs were then randomly assigned metallicities drawn from this distribution, with each member of the binary receiving the same metallicity value.

\subsubsection{Synthetic Magnitude Generation}
We assumed that \boo\ is a mono-age ancient stellar population, and adopted the 13 Gyr, solar-scaled isochrones ($[\rm{Fe}/\rm{H}] = [\rm{M}/\rm{H}]$) from the Dartmouth Stellar Evolution Database \citep{dartmouth} to transform the masses and metallicities into absolute magnitudes. This age estimate is consistent with previous investigations (e.g. \citealt{brown_sfh}), and, as noted in \cite{geha_2013}, the assumption of a single age has little effect on the determined low-mass IMF for ancient populations. The minimum metallicity available in the adopted isochrone grid is $-2.5$ dex. We used the $[\rm{M}/\rm{H}] = -2.5$ models for all stars with $[\rm{M}/\rm{H}] \le -2.5$, as there should be essentially no difference in $\rm{F606W} - \rm{F814W}$ color between stars of $[\rm{M}/\rm{H}] < -2.5$ and $[\rm{M}/\rm{H}] = -2.5$ at the same (low) mass (see Paper~I). We used a GPU-accelerated version of the \texttt{Isochrones} code (\citealt{morton_isochrones}, GPU version kindly provided by L\'aszl\'o Dobos, priv. comm.) to interpolate along the $[\rm{M}/\rm{H}]$ and mass directions to each metallicity and mass pair. The mass at the MSTO in these models is $\lesssim 0.77 \msun$, safely below the limit given above. We added the fluxes of the primary and secondary stars to determine the magnitudes of the binary pairs.

We next incorporated the distance modulus and the effects of reddening and extinction to create error-free apparent magnitudes. We allowed the distance and extinction to vary by $\pm 1$~kpc and $\pm 20\%$, respectively, by drawing from uniform distributions centered on a heliocentric distance of 65~kpc \citep{okamoto_2012} and $\rm{E}(\rm{B}-\rm{V})= 0.04$ \citep{brown_sfh} before generating each synthetic population. We assume that the reddening and extinction do not vary across the field of view.

Finally, we applied the effects of photometric errors and incompleteness to match the characteristics of the data. We then applied magnitude cuts identical to those applied to the real, observed population ($22.6~\le~F814W~\le~27.4$, and $F606W~\le~28.2$). The resulting ``observed'' synthetic photometry was then ready for comparison to the real, observed photometry, and we describe the comparison below.

\subsection{Approximate Bayesian Computation Markov Chain Monte Carlo}\label{sec:imf_fit}
\begin{figure}
\includegraphics[width=.475\textwidth]{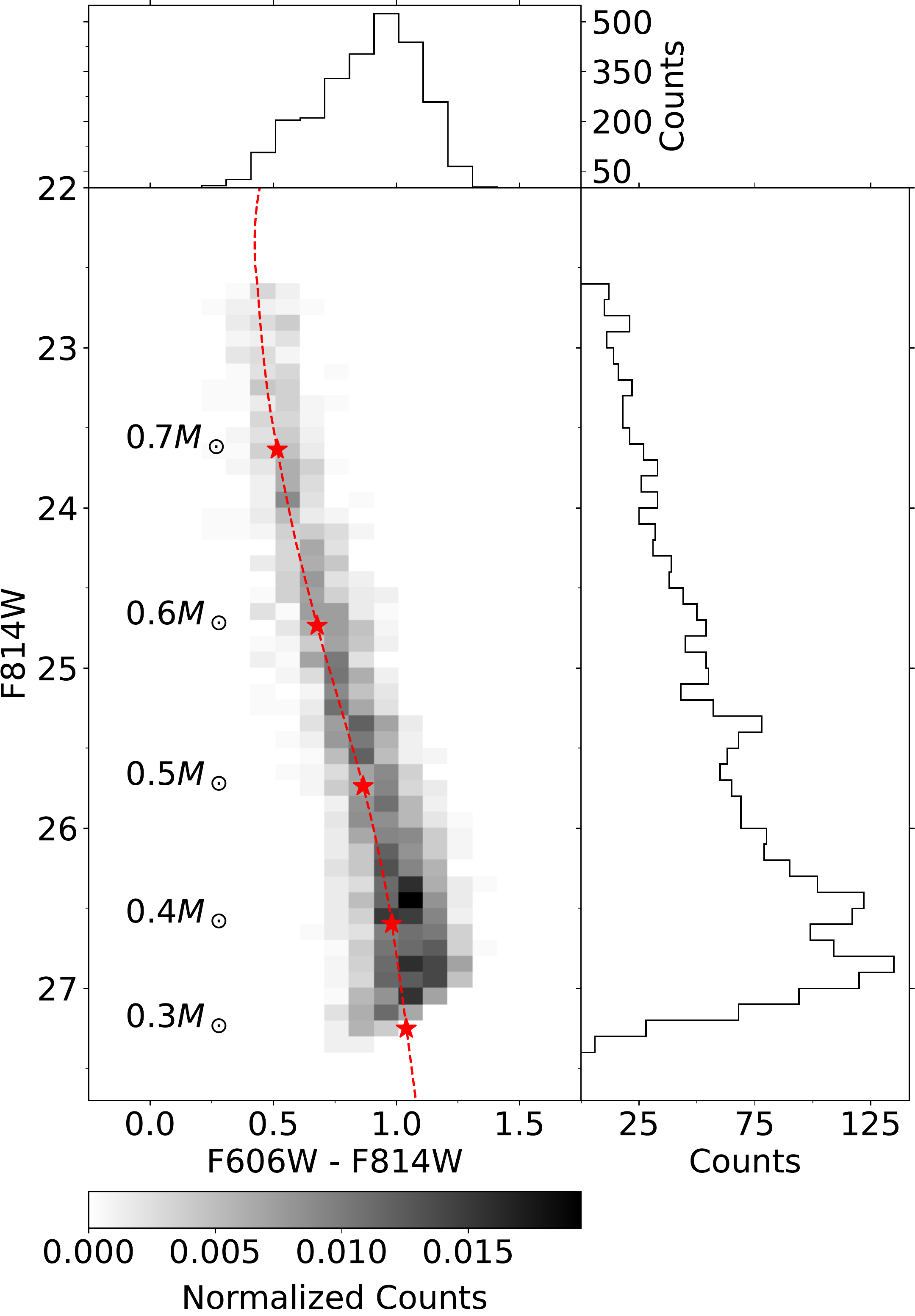}
\caption{The normalized Hess diagram of the final \boo\ catalog (left), binned identically to the procedure adopted in the ABC MCMC algorithm. To the right is the luminosity function of these data, and above is the distribution in color. Overlaid on the Hess diagram as a red dashed line is a $[\rm{M}/\rm{H}] = -2.5$ isochrone, with star-shaped points indicating the colors and magnitudes corresponding to the stellar masses given to the left of the isochrone.}

\label{fig:hess_lf}
\end{figure}

The use of an ABC MCMC algorithm to constrain the parameters of the IMF from observations is discussed at some length in \cite{gennaro_2018a}, to whom readers are referred for an in-depth discussion. We compared the two datasets using normalized Hess diagrams (such that the counts in each Hess diagram summed to unity), created by binning the photometry into $0.1$ magnitude wide square pixels over the range set by the data, namely $0.2089 \le (\rm{F606W}-\rm{F814W}) \le 1.4089$ and $22.6 \le \rm{F814W} \le 27.4$, $\rm{F606W} \le 28.2$. We present the normalized Hess diagram of the real, observed data in Figure \ref{fig:hess_lf}. The implementation of ABC requires a summary statistic that describes, through a single numerical value, how similar the simulated data are to the observed data. We adopted as the summary statistic the Jensen-Shannon distance (as implemented in \texttt{Numpy}) between the two normalized Hess diagrams in the results presented below. We verified that {similar results would have been be obtained had we adopted a different summary statistic, such as the sum of the absolute pixel-by-pixel difference between the Hess diagrams.}


There is infinitesimal probability that a synthetic dataset will be identical to the real data. Therefore, a small, non-zero threshold value for the ABC summary statistic is adopted, such that the real and synthetic data are considered to match for values below this threshold. Following \cite{gennaro_2018a}, we used an MCMC algorithm, as implemented in the \texttt{emcee} sampler \citep{emcee}, to propose IMF parameters and binary fraction values. We then generated synthetic CMDs with these parameters following the prescription given above, and we computed the summary statistic between the real and ``observed'' synthetic Hess diagrams. We began with large values of the threshold, which we then gradually reduce such that eventually the synthetic ``observed'' Hess diagrams closely matched the real one.

Specifically, we used one hundred walkers and began each experiment with fifty steps, where here an \lq experiment' refers to the ABC MCMC fit performed for a given form of the IMF. In each step, IMF parameter values and binary fractions are sampled from the priors given above (one binary fraction and set of IMF parameters for each walker). At the end of these fifty steps, we calculated the $90\%$ quantile of the summary statistics of the set of walkers, and used this value as the new threshold. We then allowed the ABC MCMC algorithm to run until all walkers had summary statistics below this threshold, and we again took the $90\%$ quantile to be the new threshold. We repeated this process until the difference between the next and previous thresholds was at most $0.5\%${, and we then ran the MCMC algorithm for an additional 3000 steps.}\footnote{For comparison, \cite{gennaro_2018a} used eighty walkers with thirty steps. They also also used the $90\%$ quantile to define the threshold, and stopped reducing the threshold once the new and old threshold are within $0.2\%$ of one another. They ran their MCMC algorithm for 2500 steps once the threshold converged}. We discarded the first 1000 steps (the \lq burn-in' period), and further thinned the resulting chains by a factor of 50. We extensively tested the ABC MCMC algorithm to ensure that it was capable of producing accurate results, as illustrated in Appendix \ref{sec:verification}. We then applied the ABC MCMC algorithm to the \boo\ member catalog.

\subsection{Results} \label{sec:results}
\begin{figure*}
\centering
\begin{minipage}{.4\textwidth}
    \centering 
    {\includegraphics[width=.75\textwidth]{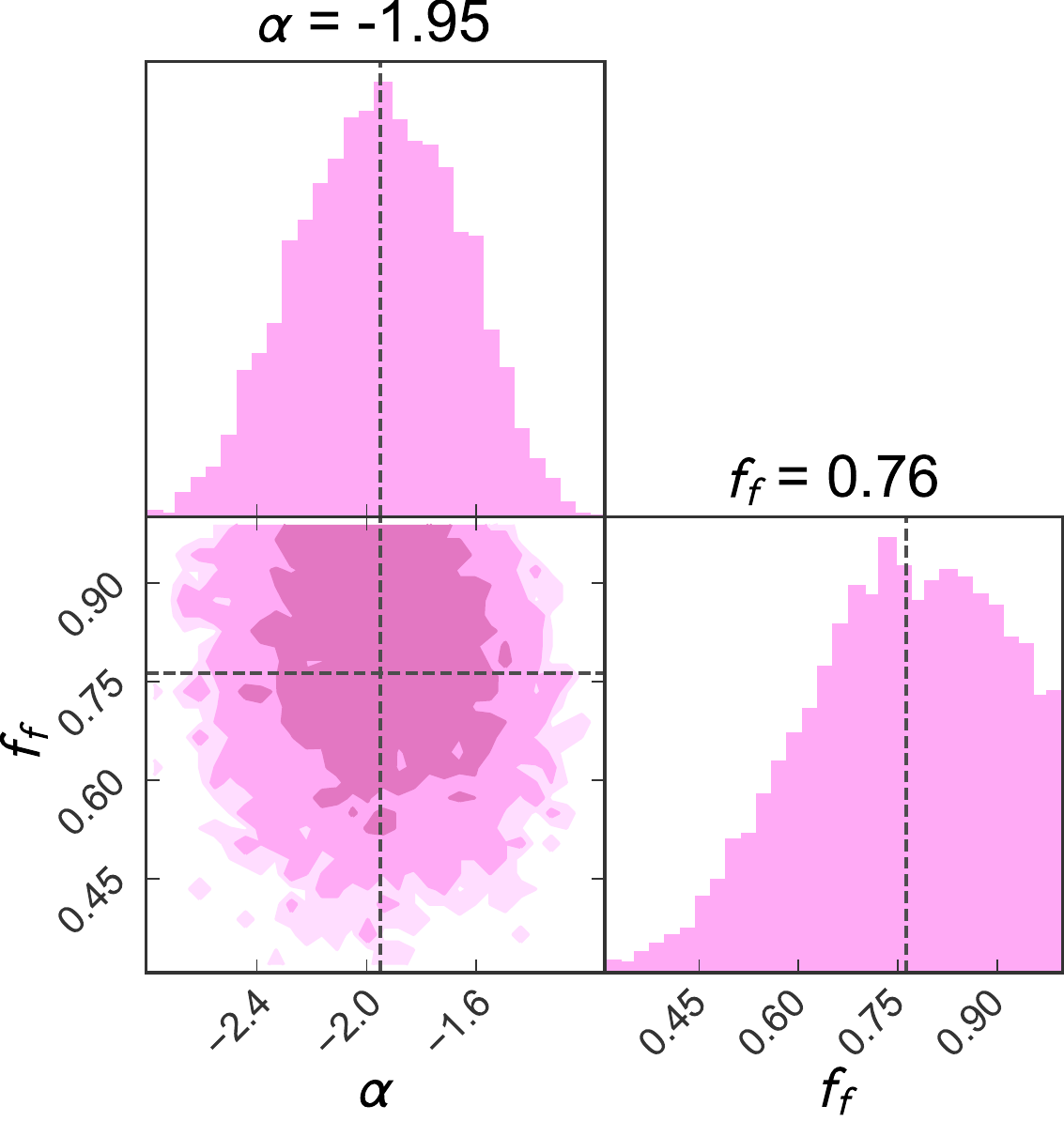}}
    \caption{The results of the single power-law fit to the \boo\ member catalog. The dashed lines on, and text above, the one-dimensional histograms indicate the best-fit values, defined as the median value of the marginalized posterior distribution. For context, the Salpeter single power-law slope is $\alpha = -2.35$.}
    \label{fig:spl_real}
    {\includegraphics[width=.75\textwidth]{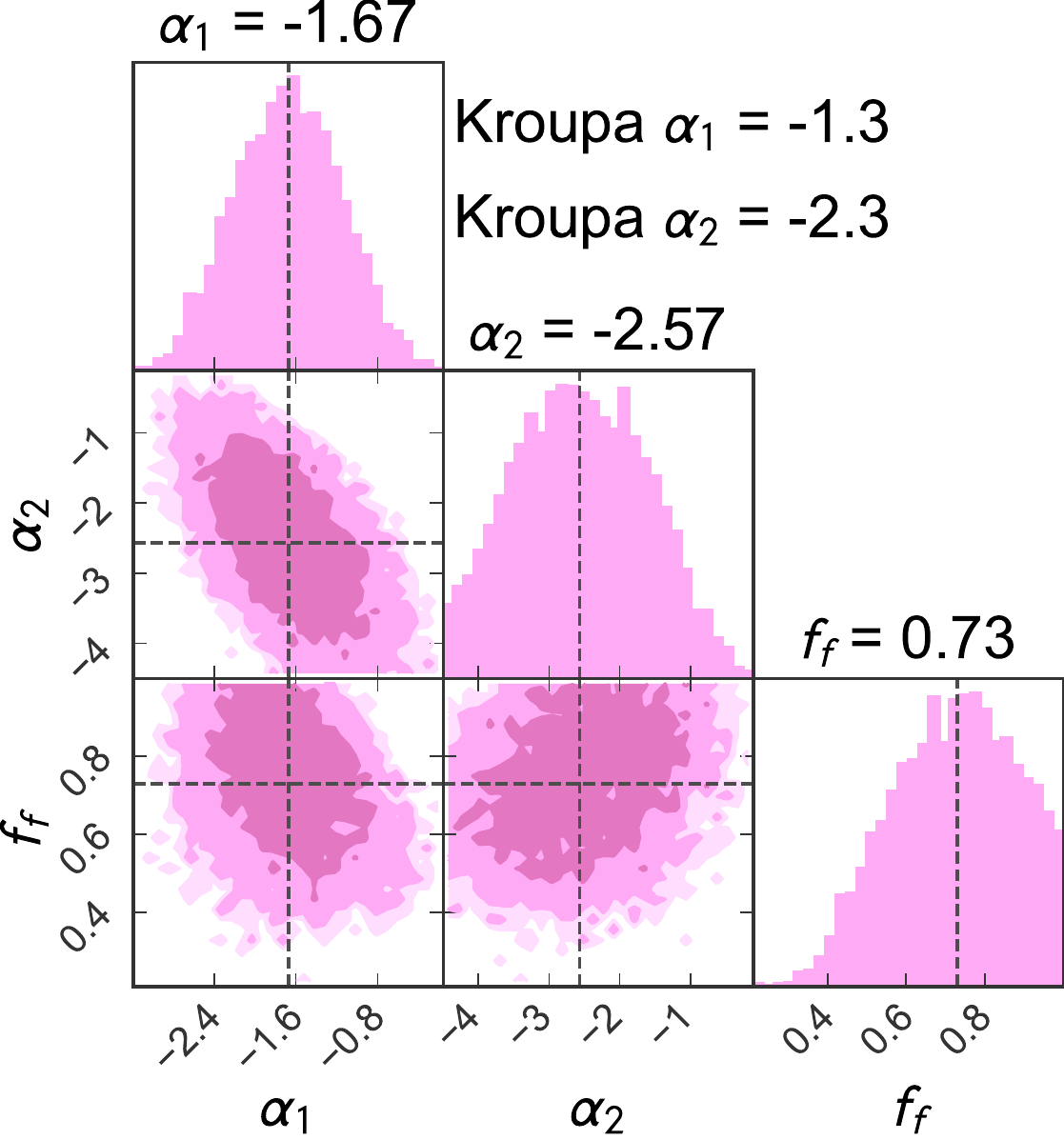}}
    \caption{The results of the broken power-law fit to the \boo\ member catalog, in which the break-mass was held fixed at $0.5\msun$. The dashed lines on, and text above, the one-dimensional histograms indicate the best-fit values, and the text to the right of the plot notes, for context, the \cite{kroupa_2001} values for the low-mass, single-star IMF of stellar populations of the Milky Way.}
    \label{fig:bpl_real}
    \end{minipage}\quad
    \begin{minipage}{.4\textwidth}
    \centering
    {\includegraphics[width=.75\textwidth]{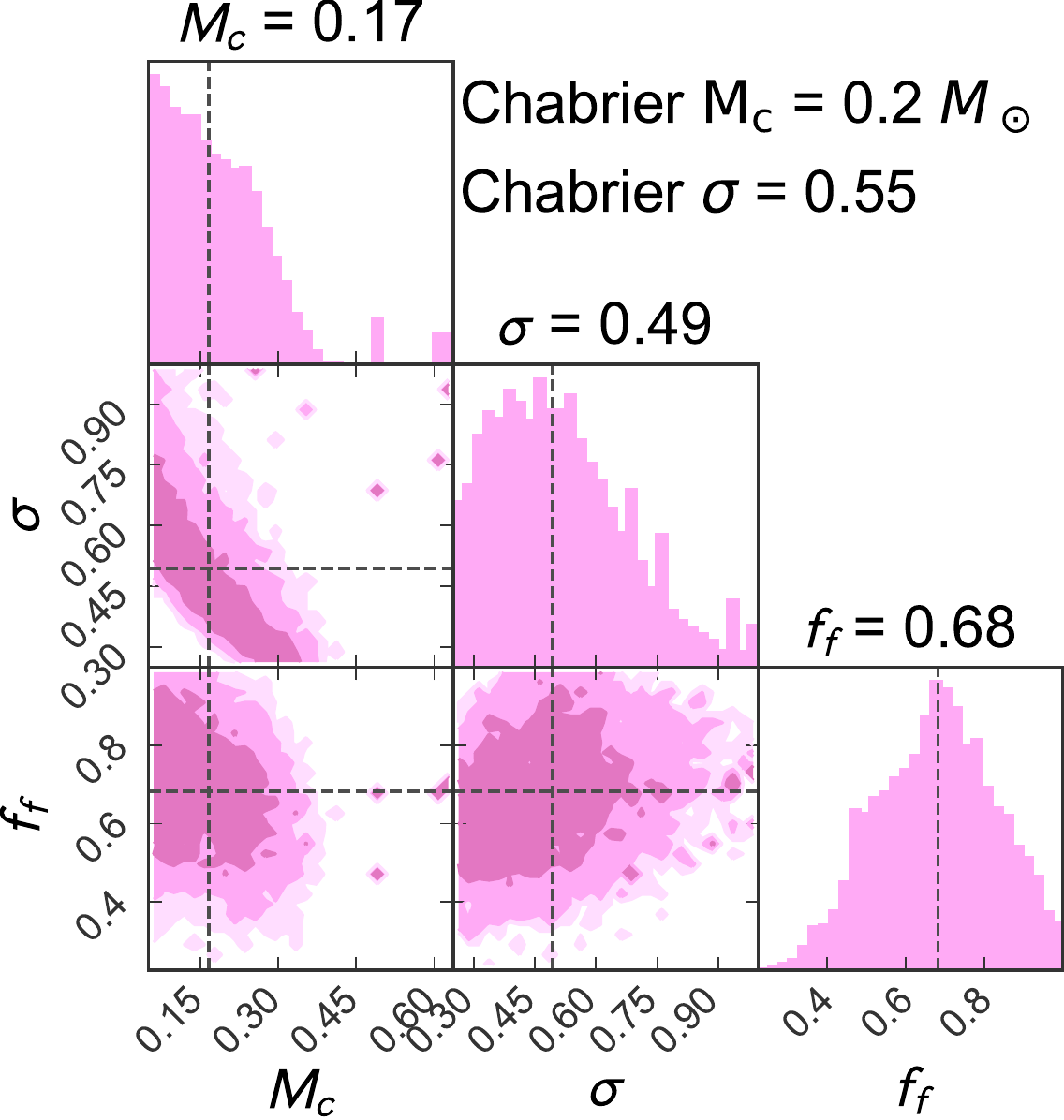}}
    \caption{The results of the lognormal fit to the \boo\ member catalog. The dashed lines on, and text above, the one-dimensional histograms give the best-fit values, and the text to the right of the plot gives, for context, the \cite{chabrier_2005} values for the low-mass, single-star IMF of stellar populations of the Milky Way.}
    \label{fig:ln_real}
    \end{minipage}\quad
\end{figure*}

The results of the ABC MCMC fits are shown graphically in Figures \ref{fig:spl_real} through \ref{fig:ln_real}, and the corresponding parameter values are given in Table \ref{tab:results}.  Figure~\ref{fig:spl_real} presents the corner plot of the results for the single power-law form of the IMF, Figure~\ref{fig:bpl_real} shows the broken power-law form, and Figure~\ref{fig:ln_real} displays the lognormal form. Each corner plot shows the pairwise correlations between the parameters as two-dimensional histograms and the marginalized posterior distributions of each parameter as one-dimensional histograms (top diagonal row). In all cases, the best-fit value is given by the median of the marginalised posterior.  We adopt the \lq highest posterior density' definition of credible interval (CI), in which the CI is the smallest interval containing the specified percentage of the posterior distribution, and we provide the $68\%$ and $95\%$ CIs of each parameter in Table \ref{tab:results}. Throughout this analysis, we adopt the $68\%$ CI as the uncertainty values on each of the presented best-fit IMF parameter values.

In each of the three IMF parameterizations, the best-fitting binary fraction (using our definition) is $f_f \sim 75\%$. As noted above in Section~\ref{sec:imf_fit}, our definition of binary fraction gives systematically higher values than the alternative definition used in the literature ($f_g$, used in e.g. \citealt{geha_2013} and \citealt{gennaro_2018a, gennaro_2018b}), and we can convert between the two definitions using the following expression: $f_g = \frac{f_f}{2 - f_f}$. Doing so, we find that the best-fit binary fraction in our definition of $f_f \sim 75\%$ translates to $f_g \sim 60\%$, which is approximately twice the best-fit value found for \boo\ by \cite{gennaro_2018a} ($f_g \sim 30\%$). If we instead take $f_f \sim 50\%$ (approximately the lower bound from the $95\%$ CIs), then the converted binary fraction is $f_g \sim 33\%$, in agreement with \cite{gennaro_2018a}.

This slight tension in retrieved binary fraction may be due to differing approaches to generating binary populations in the synthetic populations. \cite{gennaro_2018a} adopt a uniform mass ratio distribution for their binary population and fit the \textit{system} IMF, whereas we randomly pair stars from the IMF and fit the \textit{single-star} IMF. Inspection of the results of the tests presented in Appendix \ref{sec:verification} presents another possibility. Figure~\ref{fig:wrongred} shows the results of a fit where the \lq real data' were generated with distance and extinction values that differed from the synthetic data. This test indicated that an offset between the assumptions about the distance and/or extinction along the line-of-sight to \boo\ and the true, underlying distance and/or extinction may bias the ABC MCMC fits towards higher retrieved binary fractions, while still allowing the correct IMF parameters to be retrieved.

\begin{deluxetable}{cccc}
\tabletypesize{\footnotesize}
\tablecolumns{4}
\tablewidth{0pt}
\tablecaption{The best-fit (median) IMF parameters for each functional form, together with the $68\%$ and $95\%$ credible intervals (CIs) from the posterior distributions. For context, the Salpeter IMF slope is $\alpha = -2.35$, the Kroupa single-star IMF has slope values of $\alpha_{\rm{1K}} = -1.3$, $\alpha_{\rm{2K}} = -2.3$, and a break mass of $\rm{M}_{\rm{BK}} = 0.5 \msun$, and the Chabrier single-star IMF has $\rm{M}_{CC} = 0.2 \msun$ and $\sigma_{\rm{C}} = 0.55$. \label{tab:results}}
\tablehead{
\colhead{Form} \vspace{-0.1cm} & \colhead{Best-fit Value} & \colhead{$68\%$ CI} & \colhead{$95\%$ CI}}
\startdata
Single Power Law & $\alpha = -1.95$ & $_{-0.28}^{+0.32}$ & $_{-0.54}^{+0.53}$ \\[0.2cm]
&  $f_f = 0.76$ & $_{-0.12}^{+0.18}$ & $_{-0.26}^{+0.24}$ \\[0.2cm]
Broken Power Law\tablenotemark{a} & $\alpha_1 = -1.67$ & $_{-0.57}^{+0.48}$ & $_{-1.05}^{+0.91}$\\[0.2cm]
& $\alpha_2 = -2.57$ & $_{-1.04}^{+0.93}$ & $_{-1.79}^{+1.65}$\\[0.2cm]
&  $f_f = 0.73$ & $_{-0.16}^{+0.17}$  & $_{-0.27}^{+0.27}$\\[0.2cm]
Lognormal\tablenotemark{b} & $\rm{M}_c = 0.17 \msun$ & $_{-0.11}^{+0.05}$ & $_{-0.12}^{+0.18}$ \\[0.2cm]
& $\sigma = 0.49$ & $_{-0.20}^{+0.13}$ & $_{-0.24}^{+0.34}$\\[0.2cm]
&  $f_f = 0.68$ & $_{-0.17}^{+0.15}$ & $_{-0.26}^{+0.30}$\\[0.2cm]
\enddata
\tablenotetext{a}{The break mass, $\rm{M}_b$, was fixed at $\rm{M}_b = 0.5\msun$} 
\tablenotetext{b}{Note that the lower bounds of the CIs for $\rm{M}_c$ correspond to the lower limit of the prior, which was imposed to ensure that all IMFs populated CMDs within the observational limits.} 
\vspace{-0.8cm}
\end{deluxetable}

\section{Discussion}\label{sec:discussion}

\subsection{Single Power Law Initial Mass Function}\label{sec:spl}

The insight that can be gained from comparison of our result to the canonical Salpeter value for the single power-law slope is limited, as it has been established that the low-mass IMF of stellar populations in the Milky Way is not well described by the Salpeter power law, or indeed any single power law. Further, the interpretation of any individual output single power-law slope should be approached with caution: as shown in \cite{elbadry_2017}, fitting a single power-law to a population with a true IMF that is lognormal may produce an artificially shallow slope. However, some insight can still be gained from the inter-comparison of single power-law IMF slopes that were determined from fits to the data for different stellar populations (e.g. a sample of UFD galaxies). Provided that the mass range and fitting procedure are identical for each of the different populations (and the potentially differing binary populations are properly accounted for), any differences in the output power-law slopes would indicate that the underlying IMFs themselves differ, even if the true IMFs were not a single power law. We fit for a single power-law form of the IMF for consistency with previous {analyses}, and in hope of future data enabling a homogeneous analysis of the low-mass IMF of UFD galaxies down to $M \sim 0.3 \msun$.

The best-fit single power-law slope that we determined for \boo\ was $\alpha = -1.95_{-0.28}^{+0.32}$, with the Salpeter value of $\alpha = -2.35$ \citep{salpeter_1955} within the $95\%$ CI. Our relatively shallow best-fit value of the slope is similar to the results for the UFD galaxies presented in \cite{geha_2013} and \cite{gennaro_2018a}. The latter set of authors include \boo\ in their analysis, albeit over a more restricted mass range, and the best-fit single power law slope determined in that work for the \textit{system} IMF of \boo\ is $\alpha = -1.84$ (or $\alpha = -1.87$, depending on the adopted prior). The \textit{system} IMF differs from the \textit{single-star} IMF in a way that is dependent on the properties of the binary population, thus transformation between the two is far from trivial. Nonetheless, we note that both of their best-fit values for the \textit{system} single power-law IMF are within the $68\%$ CI of our fit for the \textit{single-star} single power-law IMF.

\subsection{Broken Power Law Initial Mass Function}\label{sec:bpl}

The best-fitting broken power-law slopes for \boo\ were $\alpha_1 = -1.67_{-0.57}^{+0.48}$ below the break mass (fixed to $\rm{M}_b = 0.5\msun$) and $\alpha_2 = -2.57_{-1.04}^{+0.93}$ above it. The Kroupa single-star IMF values ($\alpha_{\rm{1K}} = -1.3$, $\alpha_{\rm{2K}} = -2.3$) are both within the $68\%$ CI of our fit, and thus the \boo\ broken power-law IMF is consistent with that found for the stellar populations of the Milky Way.

Interestingly, the best-fit slopes that we determine are both steeper than the Kroupa values, which means there should be relatively \textit{more} low-mass stars. However, the uncertainties are large for both our determined slope values and the Kroupa values. For example, the uncertainty on the Kroupa $\alpha_{\rm{1K}}$ value is itself $\pm 0.5$ (given in \citealt{kroupa_2001} as the \lq $95\%$ confidence interval'). We further note that by holding the break mass value fixed, we may have underestimated the uncertainty of our fits for the other parameters, as discussed in \cite{elbadry_2017} in the context of the lognormal IMF.

\subsection{Lognormal Initial Mass Function}\label{sec:ln}
We found that the best-fitting lognormal IMF parameters for \boo\ are a characteristic mass $(\rm{M}_c)$ of $0.17_{-0.11}^{+0.05} \msun$ and a width parameter ($\sigma$) of $0.49_{-0.20}^{+0.13}$. These values are quite close to the Chabrier values ($\rm{M}_{cc}~=~0.2~\msun$, $\sigma_{\rm{C}}~=~0.55$), and the Chabrier values fall within the $68\%$ CI of the \boo\ posterior distributions. We note that we cannot strongly constrain characteristic masses below the observational limits of this study, but we can rule out that $\rm{M}_c \gtrsim 0.3\msun$ (see the discussion in \citealt{elbadry_2017}). That said, the \boo\ lognormal IMF is consistent with that of the stellar populations of the Milky Way, echoing the results above for the broken power-law IMF.

\cite{gennaro_2018a} found the best-fit \textit{system} lognormal IMF, and we did not attempt a direct comparison to our \textit{single-star} IMF, but note that their results are also consistent with those found for the \textit{system} IMF of the low-mass populations of the Milky Way ($\rm{M}_c = 0.25$ and $\sigma = 0.55$, with an assumed binary fraction for $\rm{M} < 1 \msun$ of $\sim 20\%$ to $40\%$, \citealt{chabrier_2005}).

\subsection{Comparison to Literature Analysis of Similar Depth}\label{sec:comber_comp}

We now compare our results to the only other similarly deep analysis of the low-mass IMF in an UFD galaxy - the analysis of ComBer presented in \cite{gennaro_2018b}. The data used in that analysis (deep HST WFC3 infrared imaging) reach to $M \sim 0.2\msun$, and those authors fit all of single power law, broken power law, and lognormal forms of the \textit{system} IMFs. \cite{gennaro_2018b} do not fix the break mass in the broken power-law IMF, and instead fit for it. They find that their best-fit broken power law and lognormal IMF parameters (defined to be the mean of the posterior distributions) are consistent with those of the stellar populations of the Milky Way (i.e. the Milky Way values are within the $68\%$ CIs of their posteriors, with CI defined in an identical way to this analysis), while the Salpeter slope is outside of the $95\%$ CI of their fit.  These results are broadly similar to those found in the analysis presented here for the broken power law and lognormal IMFs of \boo.

Of the set of UFD galaxies analysed in \cite{gennaro_2018a}, \boo\ and ComBer were among the least distant, and had best-fit IMF parameter values that were most consistent with the Milky Way. The analysis of \boo\ (presented here) and ComBer (presented in \citealt{gennaro_2018b}) using deeper data confirms that the IMFs of both galaxies are consistent with that of the Milky Way at the one-sigma level. As discussed in \cite{gennaro_2018a, gennaro_2018b}, \boo\ and ComBer subtend relatively large areas on the sky, which may increase the level of background galaxy contamination, particularly at fainter magnitudes, which could make the derived IMF artificially steeper and closer to the Milky Way values. Our analysis adresses this by developing a new, Bayesian approach to removing likely non-member sources (including background galaxies), and the IMF parameters retrieved for \boo\ from fits to these data are still largely consistent with those of the stellar populations of the Milky Way and the results for \boo\ presented in \cite{gennaro_2018a}. It thus seems increasingly unlikely that background galaxy contamination is biasing the fits in any off these analyses towards Milky Way values.

\subsection{Cosmological Context}\label{sec:hierarchical}
In the context of hierarchical galaxy formation, the Milky Way should have acquired a substantial fraction of present-day stellar mass from merging and assimilation of satellite galaxies. Indeed, numerous chemo-dynamical studies have concluded that a significant component  of the stellar halo is comprised of stellar debris from accreted and tidally disrupted dwarf galaxies (see e.g.~the review by \citealt{helmi_2020}). In this context, it is intriguing that the single power-law IMF that is obtained by \cite{hallakoun_2021} for the population that they identify as likely accreted stars in the halo of the Milky Way is consistent (within 2-sigma) with the single power-law IMF found here for Boo~I. \cite{hallakoun_2021} identified the blue sequence seen in the Hertzsprung-Russell diagram of stars selected to have high transverse velocities in the local {\it Gaia\/} sample as representing accreted stars. They not only found that a single power-law IMF with slope $\sim -2$ was a good fit to these stars, they also found that the blue sequence had a distinct IMF compared to both the red-sequence, high transverse-velocity stars and the low transverse-velocity populations, each of which was largely consistent with the established IMF of local stars found by \cite{kroupa_2001}.  Our result for the broken power-law form of the IMF for Boo~I is also consistent with \cite{kroupa_2001}. It is clearly necessary to have confirmation of the result for the \lq accreted stars'.  The stars in the blue sequence identified in \cite{hallakoun_2021} are a mix of high- and low-alpha populations, over a broad range of metallicities. The low-alpha population is concentrated around an iron abundance of $\sim -1.3$~dex (e.g. \citealt{hayes_2018}), also the {regime where the kinematically defined, probably accreted population dubbed   {\lq the sausage'  \citep{belokurov_2018} or \lq \textit{Gaia} Enceladus' \citep{helmi_2018} is dominant.}} Thus a selection on {alpha-to-iron abundance ratios} could potentially provide a cleaner sample of likely accreted stars. The combination of \textit{Gaia} DR3 and elemental abundance surveys have made such an IMF comparison possible (but beyond the scope of this paper).

\eject

\section{Conclusion} \label{sec:conclusion}
In this work, we used ultra-deep optical HST imaging data to constrain the low-mass, single-star IMF of \boo. These data reach to $\rm{M} \sim 0.3 \msun$, which enabled us to extend our analysis beyond the more shallow analysis of \boo\ presented in \cite{gennaro_2018a}. Using these data, we found that the Milky Way values for the broken power law and lognormal IMFs are within the $68\%$ CIs of the best-fit parameters that we determined for \boo. As such, both the broken power law and the lognormal forms of the low-mass IMF in \boo\ are consistent with those of the stellar populations of the Milky Way. This is in agreement with the results presented in \cite{gennaro_2018a} using independent, shallower data, for the \textit{system} lognormal IMF of \boo\ (where the system IMF is that inferred from star counts when binary star systems are treated as single sources). The Milky Way-like low-mass IMF of \boo\ is also in agreement with the similarly deep analysis of ComBer presented in \cite{gennaro_2018b}. The best-fit single power-law slope that we determine is formally shallower than the Salpeter slope, but this does not indicate tension with the low-mass IMF of the stellar populations of the Milky Way, as, to the best of our knowledge, no data has determined such a slope in the Milky Way below $\sim 1\msun$ (see also above discussion and \citealt{elbadry_2017}). 

With the launch of JWST, it will soon be possible to image other, more distant, UFD galaxies down to sufficiently faint magnitudes (i.e. low stellar masses) that it will be possible to  constrain robustly their low-mass IMFs. It is particularly important to obtain deeper imaging for the more distant UFD galaxies investigated in \cite{geha_2013} and \cite{gennaro_2018a}, as those authors found that the more distant galaxies tended to have the shallowest single power-law slopes. These future data will reveal whether the Milky Way-like low-mass IMFs of \boo\ and ComBer are unusual, or if in general UFD galaxy IMFs are more consistent with the stellar populations of the Milky Way than earlier indications.

{The photometric and astrometric catalogs of {sources in} the \boo\ field that we {used} in this work are made available in electronic form at the Mikulski Archive for Space Telescopes (MAST) as a High Level Science Product via \dataset[doi:10.17909/g37r-1r51]{\doi{10.17909/g37r-1r51}}. The formats of these data are given in Table \ref{tab:phot_cat} and \ref{tab:catal}, respectively.}

\section{Acknowledgements} \label{sec:acknowledgements}
{We thank the referee for their constructive comments that improved this manuscript.} CF thanks Tam\'{a}s Budav\'{a}ri, Graeme Addison, and Sam Pringle for sharing their insight and for helpful discussions. RFGW is grateful to her sister, Katherine Barber, for her support. We thank L\'aszl\'o Dobos and Jay Anderson for graciously making their software packages available to us. CF and RFGW are grateful for support through the generosity of Eric and Wendy Schmidt, by recommendation of the Schmidt Futures program. CF acknowledges support by the NASA FINESST grant (80NSSC21K2042). The authors also acknowledge grant support for HST program GO-15317, provided by NASA through grants from the Space Telescope Science Institute (STScI), which is operated  by the Association of Universities for Research in Astronomy, Inc., under NASA contract NAS~5-26555. 

Much of the data analyzed in this paper were obtained from MAST at STScI.  The specific observational datasets can be accessed via \dataset[doi:10.17909/f72q-3e91]{https://doi.org/10.17909/f72q-3e91}. Support for MAST is provided by the NASA Office of Space Science via grant NAG5–7584 and by other grants and contracts.

This work has made use of data from the European Space Agency
(ESA) mission {\it Gaia} (\url{https://www.cosmos.esa.int/gaia}),
processed by the {\it Gaia} Data Processing and Analysis Consortium (DPAC,
\url{https://www.cosmos.esa.int/web/gaia/dpac/consortium}). Funding for the DPAC has been provided by national institutions, in particular the institutions participating in the {\it Gaia} Multilateral Agreement.

\software{DrizzlePac \citep{drizzlepac}, emcee \citep{emcee}, Isochrones \citep{morton_isochrones}, matplotlib \citep{matplotlib}, numpy \citep{numpy}, pandas \citep{pandas1, pandas2}, pygtc \citep{pygtc}, scipy \citep{scipy}, TensorFlow \citep{tensorflow}}

\facility{Hubble Space Telescope}

\appendix

\section{Astrometric Data}\label{sec:astrm}

The first epoch data are comprised of deep HST ACS/WFC images of three slightly overlapping fields in the direction of \boo\ taken in 2012 as part of GO-12549 (PI: T.~Brown), in both the F606W and F814W filters. We utilized the data for only pointings 1, 3, and 5, which consist of a total of 24 images, equally divided among the three pointings, with  exposure times ranging from 430~s to 670~s. We obtained second-epoch data in 2019 for these three fields (GO-15317; PI: I.~Platais), matching the centers and orientations from GO-12549, but with significantly longer exposure times ($\sim 1240$~s). More details of the second-epoch data, including the footprints of the pointings, are presented in Paper~I. The longer exposure times were necessary to reach the required depths for our main science goal of probing the low-mass stellar mass function, although, as discussed further below, the deeper images have a somewhat detrimental effect on the astrometry of partially resolved background galaxies, due to the different surface-brightness limits reached compared to the first epoch images. Furthermore, stars bright enough to have  counterparts in the {\it Gaia\/} catalogs have saturated images in these deeper data and thus we could not obtain acceptable astrometric measurements that would have otherwise provided a direct calibration to the {\it Gaia\/} frame. 

The input images were the bias- and dark-subtracted, flat-fielded and charge-transfer inefficiency-corrected image files (i.e.~the \texttt{\_flc.fits} files). We  downloaded these for both the first-epoch and the second-epoch observations from the Mikulski Archive for Space Telescopes (MAST). We then utilized the software package \texttt{hst1pass}, which is based on the effective Point-Spread Function (ePSF) technique developed by \cite{anderson_king_2006}, which J.~Anderson kindly made available (Anderson, priv. comm.), in order to find sources in the images and calculate their precise positions in pixel coordinates, together with instrumental magnitudes\footnote{Defined by first evaluating $-2.5\log \Sigma(\rm{DN_i}\times\rm{gain})$, where ${\rm DN}_i$ represents the charge counted in each of the inner 5x5 pixels of the best-fit ePSF, and then scaling up to obtain the total charge within a radius of 10 pixels.} and a quality-of-fit parameter,\footnote{Evaluated as the sum of the absolute values of the residuals from the fit to the PSF, divided by the total charge (in electrons).} \texttt{qfit}, for each source. 

The positions in pixel coordinates are nominally precise to $\sim 0.02$~pixels, or $\sim 1$~mas (see, for example Figure~2 of \citealt{platais_2020}),  a  significantly higher level of precision than that of the FITS header WCS celestial coordinates, which is limited by the characteristics of the instrument and the guide-star acquisition. Proper motions are therefore usually  calculated in pixel-space and in the plane of gnomonic projection and we followed that practice here.

The first-epoch images produced, on average, around 6,000 detections of sources (a mix of stars, background galaxies, cosmic rays and instrumental artifacts), while the deeper second-epoch images produced about 15,000 detections. We measured positions for all the sources in ACS/WFC pixel coordinates and corrected these for geometric distortions, following the procedures{\footnote{Note that the geometric distortions for the ACS/WFC camera are both filter- and time-dependent.} given in \cite{kozhurina_2018}, with updated values provided by   Kozhurina-Platais (priv. comm., 2022), based on astrometry from {\it Gaia\/} Data Release~2
(DR~2, \citealt{gaia2}). 

The longer exposure times and fainter surface brightness detection threshold of the second-epoch images, compared to the first epoch data, meant that a given galaxy could have a measurably different shape in each set of images. This effectively rules out the template-fitting approach, previously applied in the fields of dwarf spheroidal galaxies by \cite{sohn_2013}, by which each individual galaxy is analyzed. We therefore identified significant population of background galaxies was based on their statistics, as described below, using the quality-of-fit parameter provided by the {\tt hst1pass} package, as follows.

The distribution of all the detected sources in one second-epoch image ({\tt jdir05g0q}) from pointing 3, in the plane of  \texttt{qfit} parameter {\it vs} instrumental magnitude, is shown in Figure~\ref{fig:qfit}. Two main features are evident: a cloud of sources with {\tt qfit} values in the range of $\sim 0.5$ to $\sim 1$, and a narrower streak of sources running from bottom left to top right. These two regions lie on either side of the empirical dividing line indicated by the red curve in Figure~\ref{fig:qfit} which is based on insight we gained from our earlier analyses of fields with higher stellar densities and hence a more populated stellar locus (e.g. \citealt{platais_2015}). The cloud on the left of/above the curve occupies the locus of cosmic rays, hot pixels, marginally resolved stellar binaries and background galaxies, while the sequence to the right of the curve is normally populated by stars. In rich stellar fields this sequence dies out as it approaches \texttt{qfit}~$\sim$1.0 (for example, Figure~4 of \citealt{platais_2015}). However, the sequence in Figure~\ref{fig:qfit} becomes more populated for {\tt qfit}~$\geq$0.8. This effect has also been seen in photometric studies of the Hubble Deep Fields (J. Anderson, priv. comm.) and a plausible interpretation is that it reflects a population of starlike galaxies that can dominate the faint counts in sparse stellar fields, such as that of \boo. Indeed, the analysis by \cite{bedin_2008} and our own results in Paper~I support the conclusion that barely resolved  galaxies can, and do, make it through otherwise-stringent photometric cuts that aim to isolate stars. A Bayesian framework, such as that developed in Section~\ref{sec:bayesian_cleaning} using information from HUDF, should provide better discrimination. Given this likely presence of galaxies at large values of {\tt qfit}, we rejected from further astrometric analysis all sources with \texttt{qfit}~$\gtrsim$1.0, indicated by the horizontal dashed blue line in Figure~\ref{fig:qfit}. We thus identified the sources below the blue line and to the
to the left of (or above) the red curve as galaxies and those to the right of (or below) the red curve as stars. We then applied these cuts and assignment as star or galaxy, for astrometric purposes only, to the sources of all the images for pointings 1, 3 and 5, in both epochs.

\begin{figure}
\includegraphics[width=.7\textwidth]{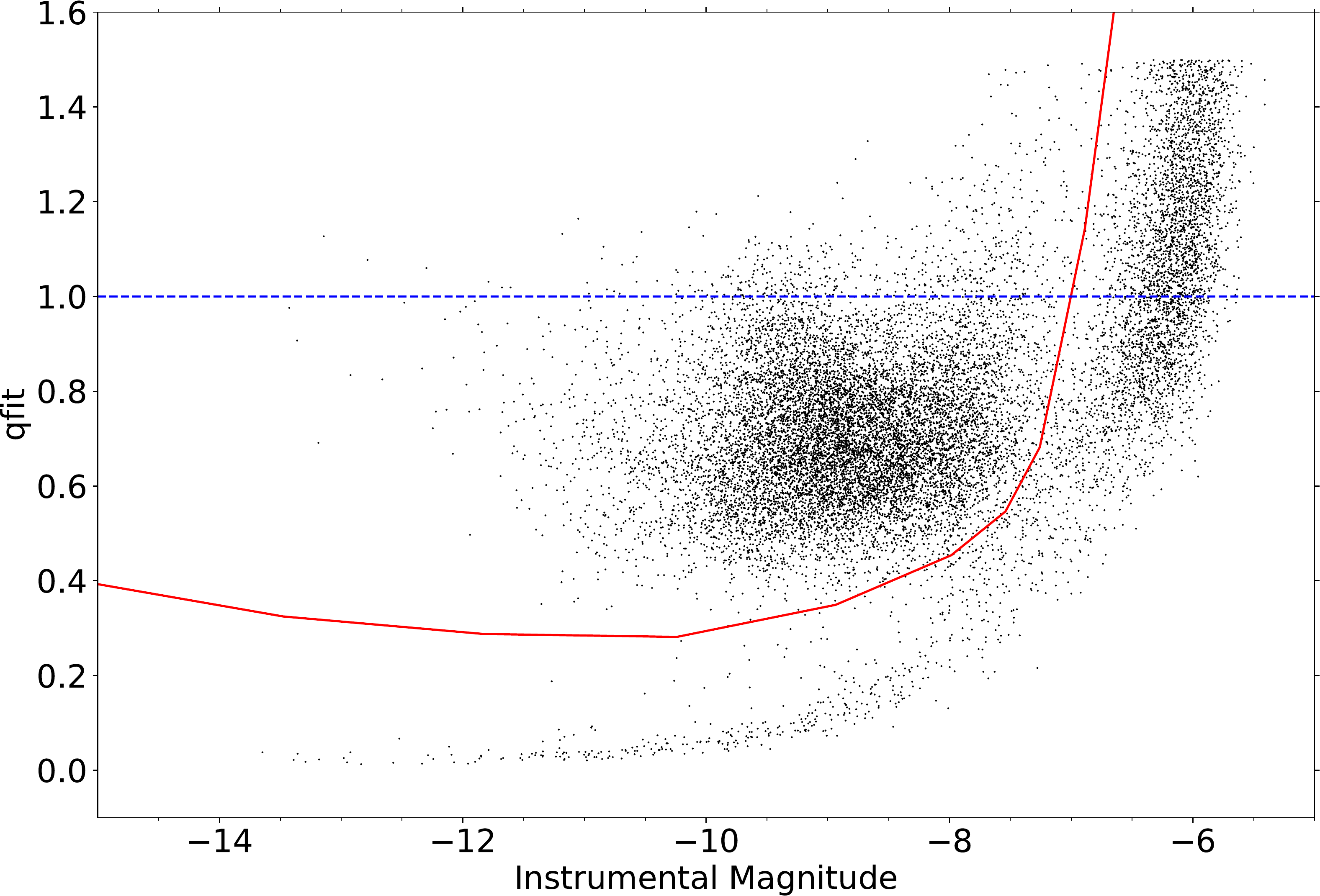}
\caption{The distribution of all sources detected in one of the second-epoch images, in the plane of instrumental magnitude {\it versus\/} the quality-of-fit parameter \texttt{qfit}.  The red curve is based upon the previous analysis of rich stellar fields and separates likely stars (to the right and below the curve) from artifacts and background galaxies (to the left and above the curve). Astrometric measurements were made only for sources with {\tt qfit} values below unity, indicated in the plot by the horizontal dashed blue line.}

\label{fig:qfit}
\end{figure}

Each pointing has  8 first-epoch and 16 second-epoch frames,  considering the F606W and F814W images together, separated by $\sim 7.1$~yr. The distortion-corrected positions in pixel-based $XY$ coordinates were grouped separately for each of the three pointings (1, 3 and 5). We then selected a ``seed'' second-epoch image for each pointing, taken through the F606W filter\footnote{The specific image files were \texttt{jdir09r2q}, \texttt{jdir05g0q} and \texttt{jdir01chq} for pointings 1, 3 and 5, respectively.} which served as the astrometric reference frame for each set of images. The transformation of all the other images (23 for each field) into the appropriate  reference frame was achieved by least-squares minimization of the differences in the positions for a set of well-measured stars (with $0<\texttt{qfit}<0.13$),  using linear and quadratic polynomial terms. The average root-mean-square error of these transformations equaled  0.022~pixel, equivalent to 1.1~mas. Since all frames were put into one of the three independent (second-epoch) reference frames (one for each field center), each object has a set of up to 24 well-aligned $XY$ positions, with the total being set by the number of images with good measurements for that object.

\subsection{Derivation of Relative Proper Motions}\label{sec:relpm}

As noted above, astrometric measurements leading to relative proper motions were obtained only for sources with {\tt qfit} values below unity, and all such sources to the right of, or below, the dividing red curve defined in Figure~\ref{fig:qfit} were identified as stars, and those to the left/above were identified as galaxies. Most of the stars are probable members of \boo\ (see Paper~I and Section~\ref{sec:membership} above) so that the registration  of all the images, including first epoch data, with the chosen second epoch reference frames means that the (member) stars will have an averaged relative proper motion of zero. It then follows that galaxies will have non-zero relative proper motions.

Each set  of aligned $X$ and $Y$ positions (in pixels) in each of the two epochs was analyzed to derive the motion on the sky for each source. {The errors in $X$ and $Y$ were estimated using the empirical relationship between instrumental magnitude and positional accuracy derived by \citet[see their Fig.~2]{bellini_2014}.} We fit a linear  trend (i.e.~a straight line) to the positions as a function of time, using a least-squares minimization. The resultant slope, plus its error, is then the relative proper motion of that source while the zeropoint of the fit provides the second-epoch coordinates (epoch 2019.5) plus associated errors. The  $XY$ coordinates and the calculated (relative) proper motion for sources in all of the three pointings were then aligned with Right Ascension and Declination axes by rotation through a fixed angle of 126.73~degree (the value being set by the choice of position angle made in the first-epoch observations, GO-12549). Note that the overlaps between adjacent pointings are very small and there were only 18 objects identified as stars in both  pointings 1 \& 3 and only 16 stars identified in both  pointings 3 \& 5. We used these stars common to two pointings in making the improvement to celestial coordinates described in Section~\ref{sec:celcoord} below.

The uncalibrated instrumental magnitude defined above (footnote eight) was sufficient to  estimate each object's centering error, which is critical in astrometric applications. However, these do not convey the faintness of the sources and we therefore defined  a \lq pseudo-Vega' magnitude  for each source, denoted by $m_{astrom}$. This was calculated by adopting the zeropoint from the ACS zero points calculator for the F775W filter (the data were taken in F606W and F814W and both filters were analysed together and  treated identically) and given by  $$m_{astrom} = -2.5 \log t_{exp} + {\rm instrumental\  magnitude} + 25.27,$$ where $t_{exp}$ is the exposure time in seconds. The resultant values agreed adequately ($\sim \pm 0.25$~mag) with the photometry in the F814W filter we derived in our earlier analysis (Paper~I).

We report proper motions only for objects that have proper-motion errors of better than 2.5~mas~yr$^{-1}$ in both coordinate axes. This cut-off yielded  measured (relative) proper motions for  a sample  of 3,634 objects identified as stars (based on their location in the plane of Figure~\ref{fig:qfit}), brighter than $m_{astrom} \sim 26.7$~mag, plus 355 objects identified as galaxies, at magnitudes brighter than $m_{astrom} \sim 25$~mag (see Figure~\ref{fig:qfit}). The astrometric catalog of these $\sim 4000$ stellar objects is publicly available online, and we outline the contents of the catalog in Table \ref{tab:catal}. The highest precision of the derived proper motions is 0.03~mas~yr$^{-1}$ for stars and 0.13~mas~yr$^{-1}$ for galaxies. 

\subsection{Interpretation of Relative Proper Motions and Absolute Proper Motion of Bo{\"o}tes~I}\label{sec:inter-relpm}

The proper-motion Vector Point Diagram (VPD) for objects brighter than $m_{astrom} \sim  25.0$~mag is shown in the leftmost panel of Figure~\ref{fig:vpd}, where it should be noted that, after the rotation described above, $\mu_{X}$ and $\mu_{Y}$ are aligned with the directions of $\mu_{\alpha\ast}$ and $\mu_{\delta}$. The vast majority of the $\sim 1000$ stellar sources, represented in the figure by filled black circles, are probable members of \boo\ (cf.~the color-magnitude diagram shown in Figures \ref{fig:star_gal_cmd} and \ref{fig:star_gal_cmd_pm}}). These cluster around the origin (zero relative proper motion). The internal line-of-sight velocity dispersion of \boo\ is $\sim 5$~km/s \citep{jenkins_2021}, giving an expected one-dimensional proper-motion dispersion of  $\sim 0.015$~mas~yr$^{-1}$ (assuming isotropic velocity dispersion tensor and a distance of 65~kpc), which is too small to be measured with these proper motions, given their errors. That said, the VPD contains only a few stars outside of the tight cluster centered on (0,0). These outliers are plausibly foreground stars in the Milky Way.


\begin{figure}
\includegraphics[width=.9\textwidth]{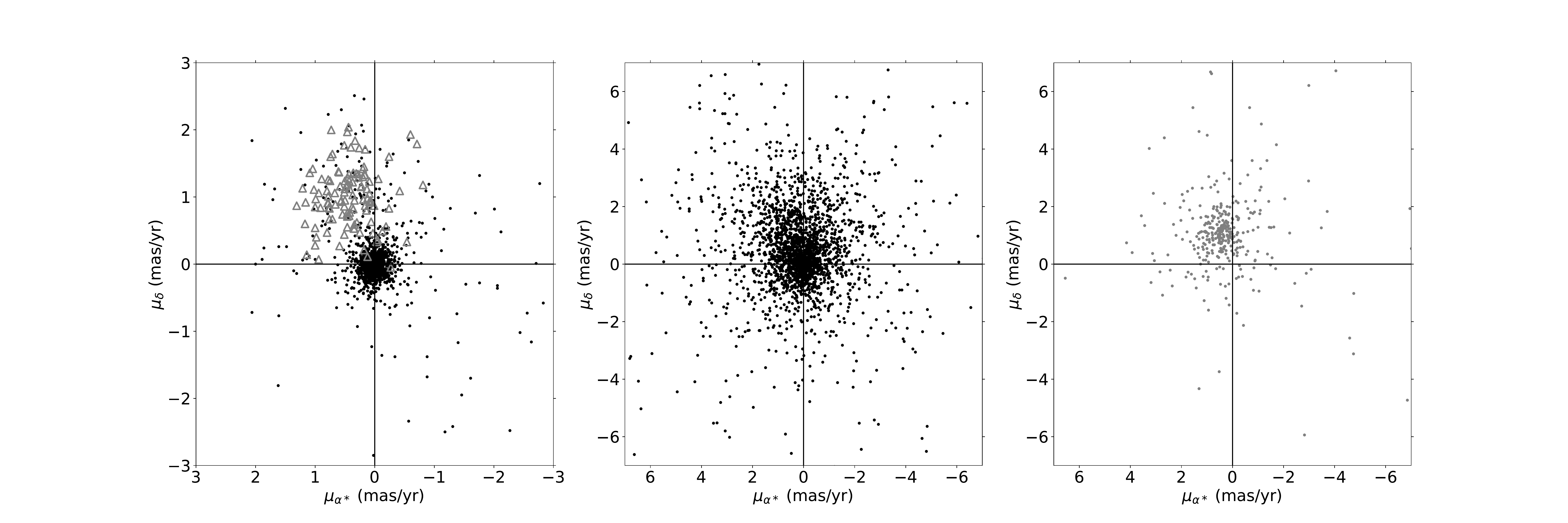}
\caption{Vector point diagrams showing the relative proper motions for sources in the \boo\ field. Left: all sources brighter than $m_{astrom} = 25.0$~mag, with stars shown as black points and the 113 best-measured background galaxies shown as grey triangles. Center:  all sources fainter than $m_{astrom} = 25.0$~mag that occupy the stellar locus defined in Figure~\ref{fig:qfit}. Right: all background galaxies (i.e. sources to the left/above the red curve in the plane of Figure~\ref{fig:qfit}) brighter than $m_{astrom} = 25$~mag. In the left panel, the stars cluster around zero, as expected, as the vast majority of these are members of \boo. The slight asymmetry in the distribution of sources in the center panel hints at the presence of faint star-like galaxies. The relative proper motions of the galaxies (right) are clearly offset from the majority of stellar relative proper motions (left).}
\label{fig:vpd}
\end{figure}
The background galaxies all have non-zero relative proper motions, representing the reflex proper motion of \boo. The grey triangles in the leftmost panel of Figure~\ref{fig:vpd} represent the subset of galaxies with the highest-precision measured relative proper motions (error less than 0.8~mas~yr$^{-1}$) and which also lie within $\sim$1~mas~yr$^{-1}$ of the peak density  of the cloud of galaxies in the VPD. This consists of a total of 113 galaxies from all three pointings. The galaxies' relative proper motions are random and therefore may be used to determine the {\it absolute\/} proper motion of \boo, by subtracting their mean value from the (zero) relative proper motion of the stars (assuming they are dominated by members of \boo). We therefore calculated the mean relative proper motion of these best-measured galaxies and estimated the associated error. We created 100,000 realizations of the galaxy proper motion data, where in each realization we drew a new proper motion value for each source from a Gaussian distribution centered on that source's measured proper motion, with width equal to the error of the measurement. We then fit a (non-binned) 2-D Gaussian to the resulting distribution of proper motions in each realization. Finally, we defined the mean reflex motion to be the median of the 100,000 2-D Gaussian means, and estimated the uncertainty to be the difference between the median and the sixteenth and eighty-forth percent quantiles, respectively. The resultant mean proper motion is then the absolute proper motion of \boo, which we determined to be $\mu_{\alpha \ast}$=0.42$\pm$0.04~mas~yr$^{-1}$ and $\mu_{\delta}$=1.00$\pm$0.04~mas~yr$^{-1}$. The estimated proper motion of \boo\ from {\it Gaia\/} EDR3 ($\mu_{\alpha \ast} = -0.397 \pm 0.019$~mas~yr$^{-1}$ and $\mu_{\delta} = -1.066 \pm 0.015$~mas~yr$^{-1}$; \citealt{filion_wyse}, see also \citealt{pace_2022} and references therein) is within 2$\sigma$ of this mean proper motion. As discussed in Section~\ref{sec:pm_boo}, this is an independent measurement that is complementary to \textit{Gaia}.

It should be acknowledged that galaxies could be sufficiently star-like to occupy the stellar locus of Figure~\ref{fig:qfit}. This is illustrated in the middle panel of Figure~\ref{fig:vpd}, which shows the proper motions for all the faint objects in the stellar locus only. Note that this plot is zoomed out compared to the leftmost panel of Figure~\ref{fig:vpd}, with both axis scales changed by a factor of $\sim 2$. The asymmetry to the top-left quadrant suggests contamination by galaxies, supported by the VPD in the rightmost panel of Figure~\ref{fig:vpd}, which shows all the 355 objects brighter than $m_{astrm} = 25$ that occupy the galaxy locus of Figure~\ref{fig:qfit}. The contaminating galaxies in the middle panel cannot be identified individually, as the errors in the measured  proper motions at faint magnitudes ($m_{astrom} > 25$~mag)  are too high ($\sim 2$~mas~yr$^{-1}$). However, we can estimate the relative contribution of these galaxies to the VPD via simple counts of sources in different locations. As illustrated in Figure~\ref{fig:vpd}, galaxies predominantly occupy the top left quadrant of the VPD, and thus contamination by these sources would lead to a relative over-density of sources in this quadrant (assuming that the near-zero relative proper motion \boo\ stars are approximately evenly distributed over the quadrants). For stars brighter than $m_{astrm} = 25$ (black points in the leftmost panel in Figure~\ref{fig:vpd}) there is no clear signal of galaxy contamination (each quadrant contains $25 \pm 3 \%$ of the sources). However, stars fainter than $m_{astrm} = 25$ (black points in the middle panel in Figure~\ref{fig:vpd}), show an asymmetry, with the top left quadrant containing $\lesssim 40\%$, indicating the likely presence of galaxy contamination.

The derivation of proper motions should allow us to detect and then remove Milky Way stars and galaxies from the deep color-magnitude data of \boo. The  astrometric catalog and the photometric catalog have rather different selection functions which argues against attempting  a statistical subtraction of galaxies from the faint star counts in the photometric catalog based on the estimates of contamination in the astrometric data described above. The difference in the selection functions is especially obvious at the faint end, as the photometric catalog reaches fainter than the the astrometric data. The proper motions for individual sources can still be used to identify and remove the (brighter) likely Milky Way contaminants, as described in Section~\ref{sec:pm_identification}.

\subsection{Improving Celestial Coordinates}\label{sec:celcoord}

We noted above, in Section~\ref{sec:astrm}, that the WCS parameters may not provide optimal celestial coordinates. Generally, celestial coordinates for the detected sources in a given field may be calculated through the identification of astrometric standard stars in that field. The very limited areal coverage of the second-epoch \boo\ observations, combined with the long exposure times (and hence saturated images of bright stars), meant that there is no  counterpart from  {\it Gaia} EDR3 in  the astrometric catalog.  We therefore used the shortest-exposure first-epoch data (430s through filter F814W) to identify \textit{Gaia} EDR3 stars. We found $\sim 15$ objects in each of the three pointings (a minimum of 11 and a maximum of 20) with WCS-based celestial coordinates that matched an entry in  {\it Gaia\/} EDR3 (adopting a matching radius of 0.8~arcsec). The largest median offset between the WCS-based coordinates and {\it Gaia\/} EDR3 was $\sim 0.5$~arcsec (for pointing~5) and the internal scatter for all pointings was $\sim 0.1$~arcsec. There is no systematic offset in RA or Dec between the first and second epochs, so that these estimated offsets are valid for data from either epoch.

Once the first-epoch celestial coordinates had been corrected using the derived offsets from the brighter stars, we used the fainter stars in common with the second-epoch data as reference stars 
to calculate a final, corrected celestial coordinates (Right Ascension and Declination, equinox J2000, epoch 2019.5)  for all the stars in the astrometric catalog.   We also applied the corrections to the coordinates for all stars for which there is  DAOPHOT photometry (obtained as described in Paper~I).   

Stars with positions in the narrow overlap areas between the independent pointings (1, 3) and (3, 5) allowed an external check of these corrections. The 18 stars in common between pointings 1 and 3 had a maximum offset of 35~mas while the 16 stars in common between pointings 3 and 5 had a  maximum offset of 50~mas. This established the accuracy of the derived celestial
coordinates.

\section{Determination of the Initial Mass Function using Alternate Binary Mass-Ratio Distributions} \label{sec:mass_ratio}

{We adopted alternate binary mass-ratio distributions to determine whether or not the choice that we made - to randomly pair masses in binaries - significantly impacted our results. We implemented both an approximately uniform mass-ratio distribution and a power-law distribution with index $\gamma = 1.4$ (with $\gamma = 1$ corresponding to the uniform mass-ratio distribution). This value of the power-law index ($\gamma = 1.4$) is generally consistent with the mass-ratio distribution of low-mass, Population I stars in the Milky Way (see e.g. \citealt{duchene_2013}). We ran each test on the actual, observed photometric data, and did not change any other aspect of the ABC MCMC algorithm (as described in the main text).}

{As in the main text, we first drew $N$ masses from the chosen IMF and assigned each mass to a \lq star'. We then randomly selected  $\frac{N f_f}{2}$ of these \lq stars' to be members of binary systems (rather than the fraction $N f_f$ used in the analysis in the main text), where $f_f$ is the binary fraction, defined (in the main text) so that  $\frac{N f_f}{2}$ is the number of unresolved binary pairs.  We then drew  a matching number of mass-ratio values from the chosen mass-ratio distribution (uniform or power-law). Each distribution was bounded between the values of $0.11$ and $1$, which correspond to, respectively, the minimum mass ratio possible given the minimum   ($0.11 \msun$)  and maximum ($1 \msun$) stellar masses, and an equal-mass binary system. We then randomly assigned each \lq star' with one of the mass-ratio values drawn from the distribution and identified the appropriate primary or secondary \lq star' from the remaining set (i.e. the \lq stars' not within the initial $\frac{N f_f}{2}$ subset). We then removed this best-matching primary or secondary \lq star' from the pool of \lq stars' drawn from the IMF, in order  to preserve the IMF and avoid double-counting. This approach slightly skews the implemented mass-ratio distribution (determined from the paired stars) away from the intended uniform or power-law distribution. This is most easily seen in the case of the uniform distribution, where the implemented mass-ratio distribution of the paired stars is skewed towards higher mass-ratio values. This bias against low mass-ratio values is unsurprising, given the stellar mass bounds: \lq stars' with mass less than $1 \msun$ cannot be in a binary pair with a mass ratio of $0.11$, and thus such a \lq star' will be paired into a binary system with the smallest mass ratio possible within the constraints.}

\begin{figure}
\includegraphics[width=.5\textwidth]{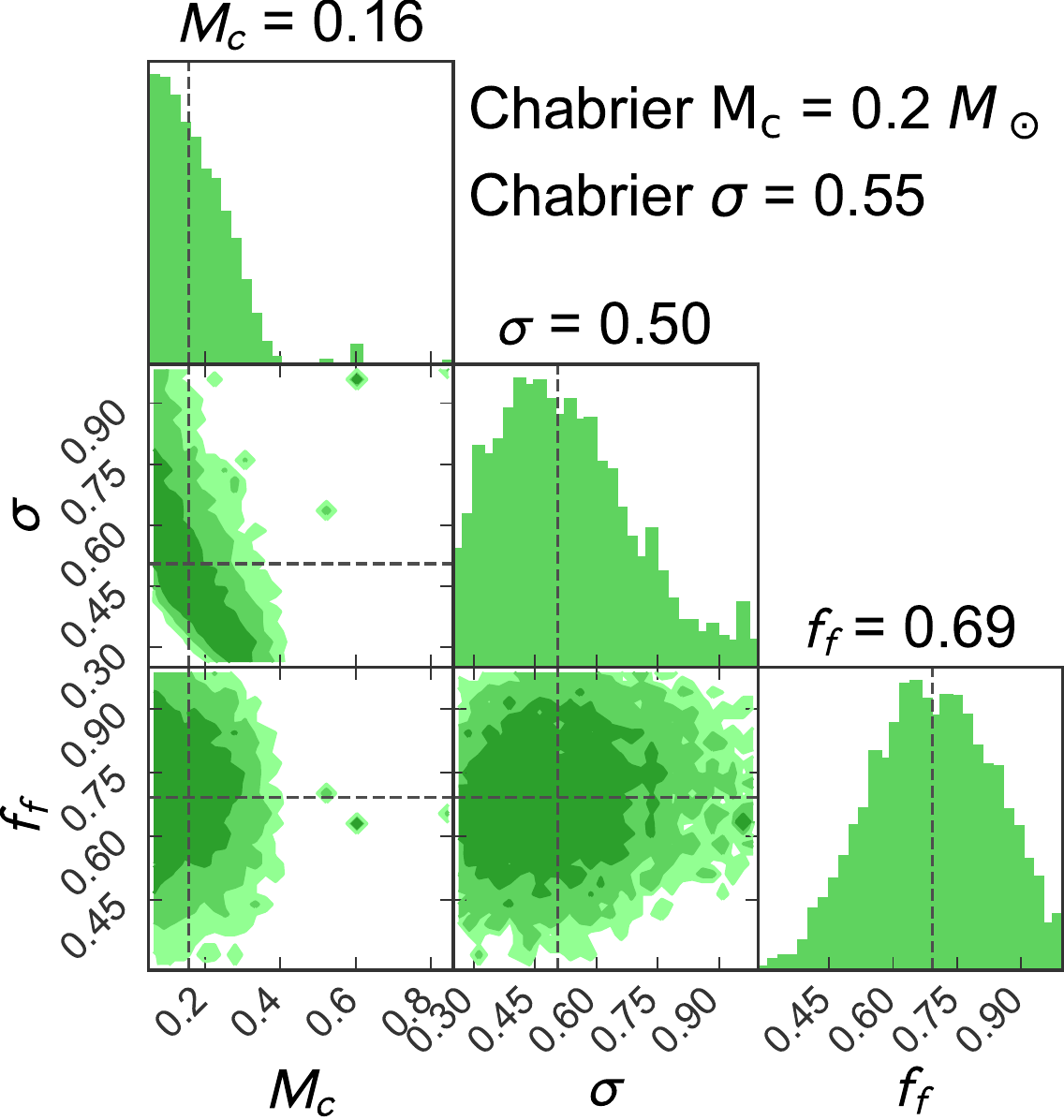}
\caption{A sample corner plot investigating the effect of assuming a uniform binary mass-ratio distribution while fitting a lognormal IMF to the \boo\ observational data. The retrieved best-fit values are given above the one-dimensional histograms (also shown with black dashed lines), and to the right, the \cite{chabrier_2005} values for the low-mass single-star IMF of stellar populations of the Milky Way are provided for context.}
\label{fig:ln_unibin}
\end{figure}

{We then ran the ABC-MCMC with each of these alternate binary mass-ratio distributions to determine their effect on the retrieved best-fit values for the broken power-law and lognormal forms of the IMF, finding little difference. We provide a summary of the results in Table \ref{tab:alt_bin}, and briefly discuss these results here. For the broken power-law IMF, we find that the best-fit values for both the uniform and power-law binary mass-ratio distributions are within the $68\%$ CI of those determined under the assumption of random mass pairing (namely $\alpha_1 = -1.67_{-0.57}^{+0.48}$, $\alpha_2 = -2.57_{-1.04}^{+0.93}$, and $f_f = 0.73_{+0.17}^{-0.16}$). The Kroupa single-star IMF values ($\alpha_{1K} = -1.3$, $\alpha_{2K} = -2.3$) are within the $68\%$ CI of the best-fit values determined for both the uniform and power-law binary mass-ratio distribution, though we note that the lower bound of the $68\%$ CI of the $\alpha_1$ fit exactly equals the Kroupa value for both cases. For the lognormal IMF, the best-fit values for uniform and power-law binary mass-ratio distributions are indistinguishable from the best-fit values obtained in Section \ref{sec:results} (i.e. a characteristic mass of $\rm{M}_c = 0.17_{-0.11}^{+0.05} \msun$, a width parameter of $\sigma = 0.49_{-0.20}^{+0.13}$, and a binary fraction of $f_f = 0.68_{-0.17}^{+0.15}$). The Chabrier single-star IMF values are within the $68\%$ CI of the fits for both the uniform and power-law mass-ratio distributions. We thus conclude that the choice of binary mass-ratio distribution does not have a significant effect on the determined best-fit IMF values. We illustrate this conclusion in Figure \ref{fig:ln_unibin}, which shows a sample corner plot of the lognormal IMF fit to the observed data if a uniform binary mass-ratio distribution were adopted. The best-fit values (defined by the median of the distribution) are given above each one-dimensional histogram and indicated with black dotted lines.}

\begin{deluxetable}{cccc}
\tabletypesize{\footnotesize}
\tablecolumns{4}
\tablewidth{0pt}
\tablecaption{The best-fit (median) IMF parameters for the broken power-law and lognormal forms of the IMF determined with either uniform or power-law binary mass-ratio distributions, together with the $68\%$ and $95\%$ credible intervals (CIs) from the posterior distributions. \label{tab:alt_bin}}
\tablehead{
\colhead{Form} \vspace{-0.1cm} & \colhead{Best-fit Value} & \colhead{$68\%$ CI} & \colhead{$95\%$ CI}}
\startdata
Broken Power Law\tablenotemark{a}, Uniform & $\alpha_1 = -1.84$ & $_{-0.58}^{+0.54}$ & $_{-0.95}^{+1.05}$\\[0.2cm]
& $\alpha_2 = -2.39$ & $_{-0.94}^{+0.97}$ & $_{-1.67}^{+1.76}$\\[0.2cm]
&  $f_f = 0.68$ & $_{-0.18}^{+0.15}$  & $_{-0.26}^{+0.31}$\\[0.2cm]
Broken Power Law\tablenotemark{a}, Power Law  & $\alpha_1 = -1.86$ & $_{-0.54}^{+0.56}$ & $_{-1.02}^{+1.00}$\\[0.2cm]
& $\alpha_2 = -2.35$ & $_{-0.98}^{+0.93}$ & $_{-1.75}^{+1.65}$\\[0.2cm]
&  $f_f = 0.62$ & $_{-0.16}^{+0.13}$  & $_{-0.25}^{+0.33}$\\[0.2cm]
Lognormal\tablenotemark{b}, Uniform & $\rm{M}_c = 0.16 \msun$ & $_{-0.11}^{+0.05}$ & $_{-0.11}^{+0.16}$ \\[0.2cm]
& $\sigma = 0.50$ & $_{-0.20}^{+0.11}$ & $_{-0.25}^{+0.33}$\\[0.2cm]
&  $f_f = 0.69$ & $_{-0.14}^{+0.15}$ & $_{-0.27}^{+0.26}$\\[0.2cm]
Lognormal\tablenotemark{b}, Power Law & $\rm{M}_c = 0.17 \msun$ & $_{-0.11}^{+0.05}$ & $_{-0.11}^{+0.19}$ \\[0.2cm]
& $\sigma = 0.52$ & $_{-0.23}^{+0.12}$ & $_{-0.25}^{+0.39}$\\[0.2cm]
&  $f_f = 0.63$ & $_{-0.11}^{+0.15}$ & $_{-0.25}^{+0.22}$\\[0.2cm]
\enddata
\tablenotetext{a}{The break mass, $\rm{M}_b$, was fixed at $\rm{M}_b = 0.5\msun$} 
\tablenotetext{b}{Note that the lower bounds of the CIs for $\rm{M}_c$ correspond to the lower limit of the prior, which was imposed to ensure that all IMFs populated CMDs within the observational limits.} 
\vspace{-0.8cm}
\end{deluxetable}

\section{Verification of the Approximate Bayesian Computation Markov Chain Monte Carlo} \label{sec:verification}

We tested the ABC MCMC algorithm on simulated populations with known IMFs and binary fractions to quantify its ability to return accurate parameter estimates. We simulated synthetic populations with each of the three forms of the IMF constrained in this work (single power law, broken power law, and lognormal), and a range of binary fractions. We required that each population have approximately the same number of sources on the \lq observed' CMD as in the real, observed \boo\ data within the imposed magnitude limits. We present the results of some of these tests here.

\begin{figure}
\centering
\begin{minipage}{.4\textwidth}
    \centering 
    {\includegraphics[width=.75\textwidth]{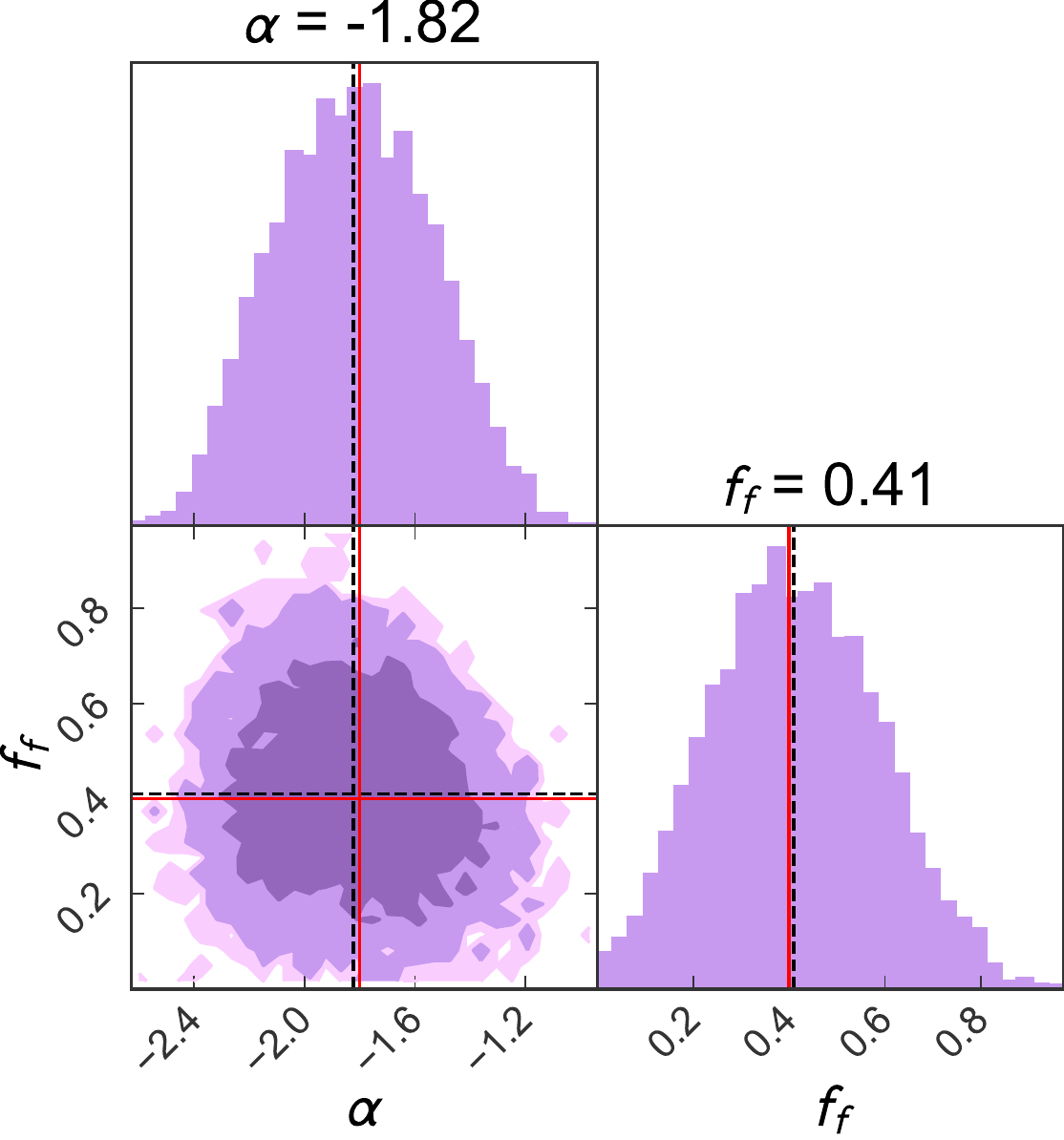}}
    \caption{A sample corner plot verifying the performance of the ABC MCMC algorithm for a simulated population with a known single power-law IMF slope and binary fraction. The {input parameter} values ($\alpha = -1.8$, $f_f = 0.4$) are shown in red, and the retrieved best-fit values are given above the one-dimensional histograms (also shown with black dashed lines).}
    \label{fig:spl_example}
    {\includegraphics[width=.75\textwidth]{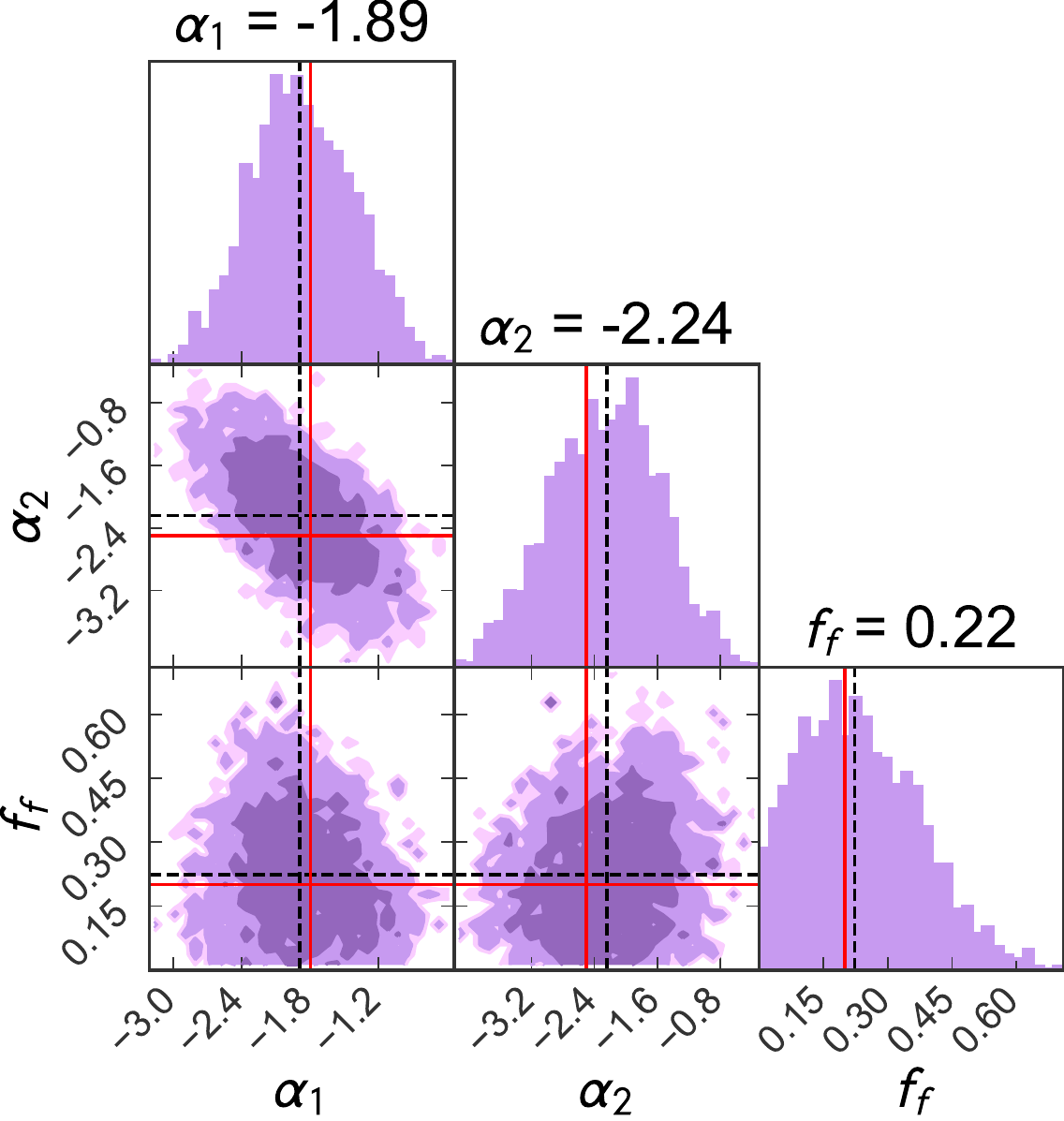}}
    \caption{A sample corner plot verifying the ABC MCMC algorithm for a simulated population drawn from a broken power-law IMF. {Input parameter values} (red) are $\alpha_1 = -1.8$, $\alpha_2 = -2.5$ and $f_f = 0.2$, and the retrieved best-fit values are given above the one-dimensional histograms (also shown with black dashed lines). As in the analysis of the observed data, the break-mass is held fixed at $\rm{M_b} = 0.5\msun$.}
    \label{fig:bpl_example}
    \end{minipage}\quad
    \begin{minipage}{.4\textwidth}
    \centering
    {\includegraphics[width=.75\textwidth]{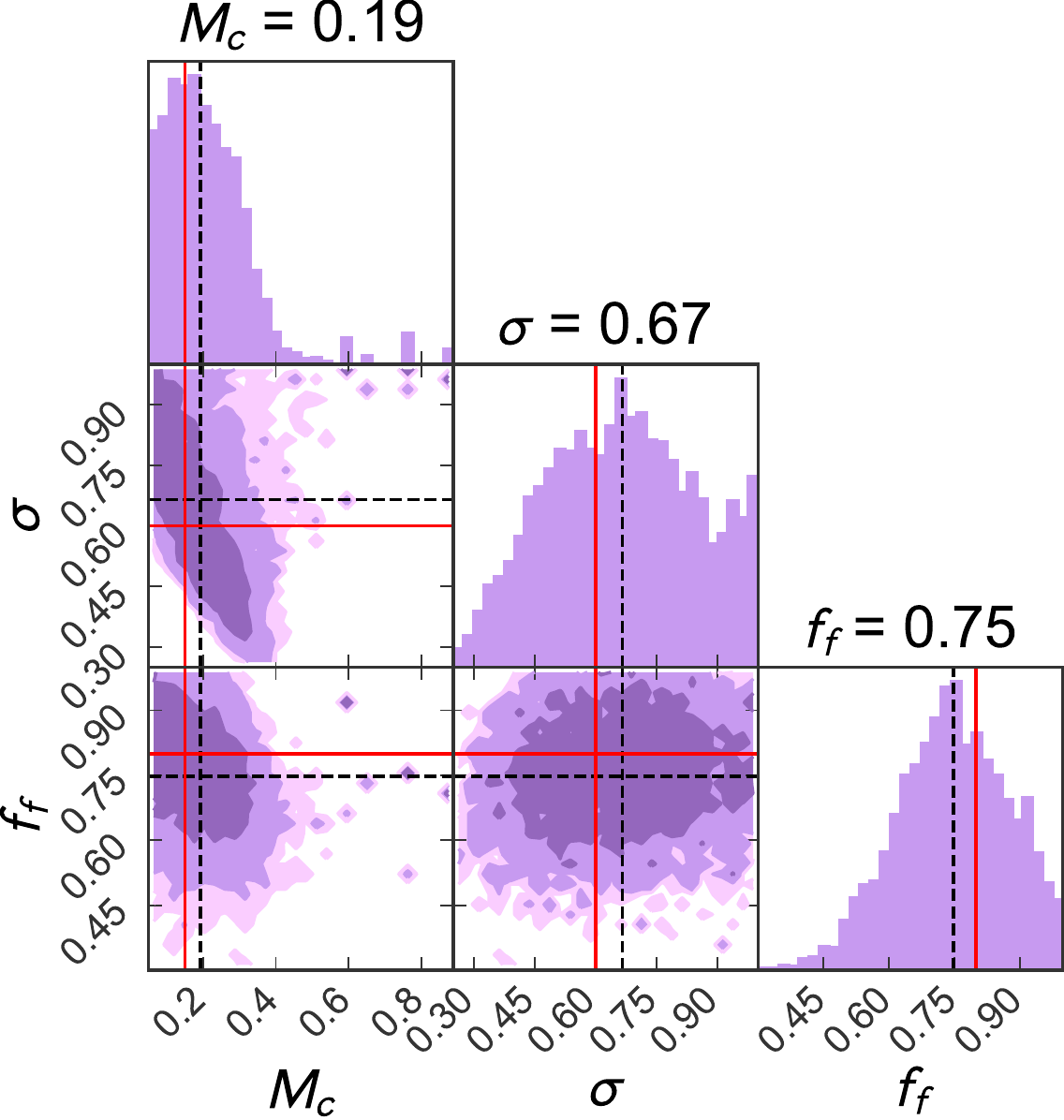}}
    \caption{A sample corner plot verifying the ABC MCMC algorithm for a simulated population drawn from a lognormal IMF.  {Input parameter} values (in red) are $\rm{M}_c= 0.15 \msun$, $\sigma = 0.6$ and $f_f = 0.8$, and the retrieved best-fit values are given above the one-dimensional histograms (also shown with black dashed lines). }
    \label{fig:ln_example}
    \end{minipage}\quad
\end{figure}

Figures \ref{fig:spl_example}, \ref{fig:bpl_example}, and \ref{fig:ln_example} show the corner plots of the fits to single power law, broken power law, and lognormal synthetic data, respectively. The black dotted lines and quantities given above each one-dimensional histogram (showing the marginal distribution) indicate the best-fit value, defined by the median of the distribution. {The input parameters are shown with red lines.} In all cases, the best-fit value was close to the {input value}, and we were able to retrieve the input values well within the $68\%$ CI. Of all of the fits, the retrieved $\alpha_2$ slope parameter in the broken power-law IMF was the furthest from the input value (as seen in Figure~\ref{fig:bpl_example}). However, $\alpha_2$ is relatively poorly constrained by the ABC MCMC algorithm, as evidenced by the large width of the marginal distribution. The $68\%$ and $95\%$ CIs are thus large, and the true value is easily retrieved within the CIs despite the deviation between the best-fit and the {input parameters}.

Finally, we verified that the IMF parameters resulting from the ABC MCMC fit were not sensitive to the exact choice of distance and extinction of the synthetic population (within the errors of those determined observationally for \boo). We generated a synthetic population with a broken power-law IMF and the same assumed population parameters as described in Section~\ref{sec:modelling}, but adopted a heliocentric distance of $63$~kpc and an $\rm{E(B-V)}$ value that was fifteen percent higher than the \cite{brown_sfh} value (compared to $65$~kpc and $\rm{E(B-V)} = 0.04$). Note that in generating synthetic populations within the ABC MCMC algorithm, we allow distance and extinction to vary by $\pm1$ kpc and $\pm 20\%$, respectively.

\begin{figure}
\includegraphics[width=.5\textwidth]{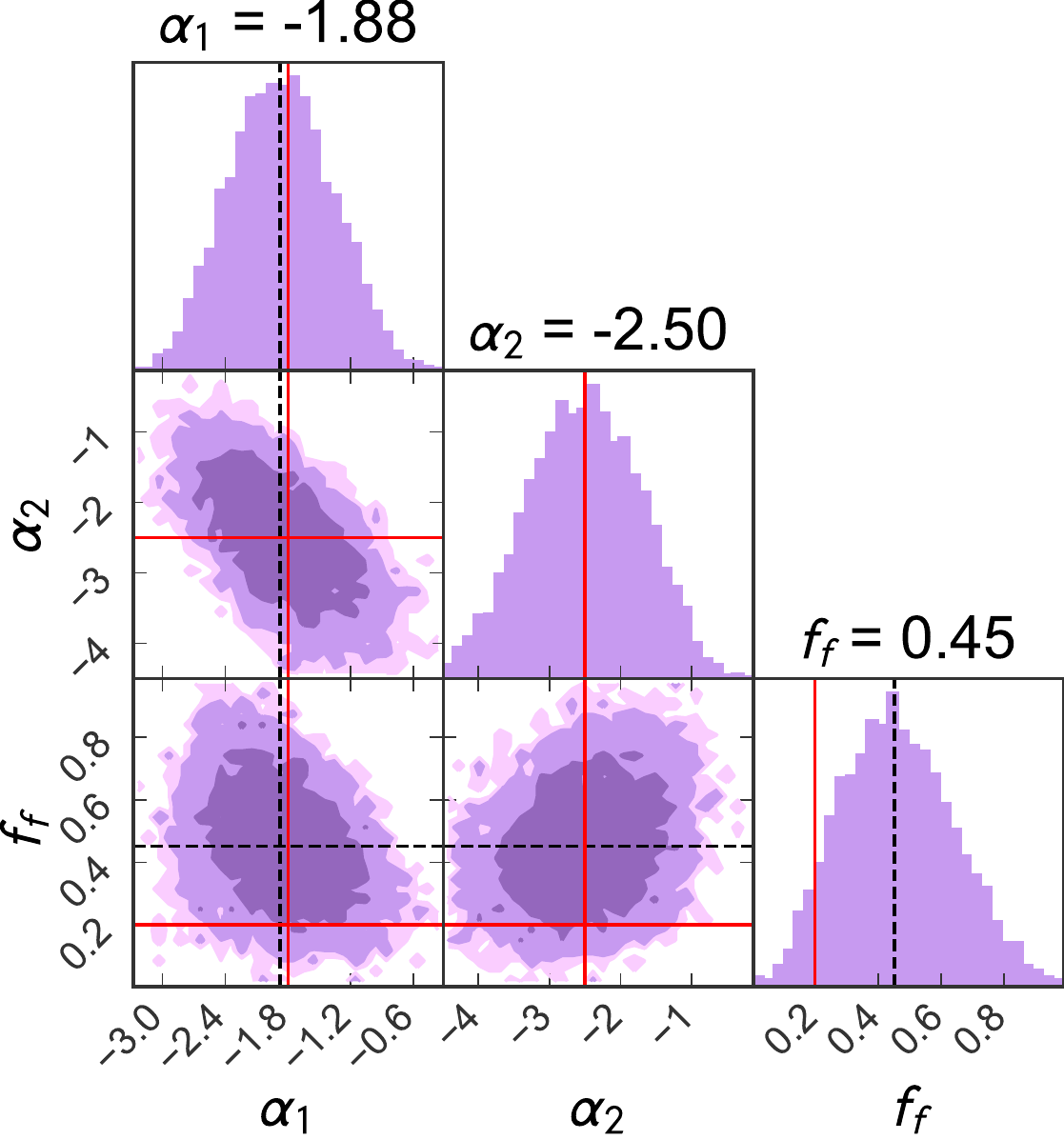}
\caption{A sample corner plot verifying the ABC MCMC algorithm for a simulated population drawn from a broken power-law IMF. The synthetic population that was used as the \lq observed data' was simulated using values for the extinction and distance that do not match the synthetic data generated in the ABC MCMC algorithm (15$\%$ higher extinction and $\sim 3\%$ closer distance). The {input parameter} values (red) are $\alpha_1 = -1.8$, $\alpha_2 = -2.5$ and $f_f = 0.2$, and the retrieved best-fit values are given above the one-dimensional histograms (also shown with black dashed lines). Note that we cannot retrieve the binary fraction within the $68\%$ CI.}
\label{fig:wrongred}
\end{figure}

We present a corner plot of the resulting ABC MCMC fits in Figure~\ref{fig:wrongred}. The retrieved slope values were quite close to the {input parameters}, but the retrieved binary fraction value was more than double the input value. The {input value} for the binary fraction is within the $95\%$ CI. {Additional tests using an adopted heliocentric distance of $67$~kpc and the same, fifteen percent higher $\rm{E(B-V)}$ value gave similar results: the retrieved IMF slope values were close to the input values, while the retrieved binary fraction was \textit{smaller} than the input. These tests indicate} that ABC MCMC algorithm is robust against (reasonable) uncertainties in the distance and extinction when retrieving the IMF parameters, but {less so} in its ability to retrieve the binary fraction. The {true} binary fraction is difficult to constrain observationally and acts as a factor that widens the observed {main sequence on the CMD, so it is perhaps unsurprising that this aspect of the fit would be poor}. We further note that when determining the best-fit values for IMF parameters (in each of the functional forms), we marginalized the probabilities over the binary fraction. This marginalization {allowed} the CIs of the IMF parameters to incorporate realistic uncertainties in the binary fraction.

\clearpage
\begin{deluxetable}{lcl}
 \tabletypesize{\footnotesize}
 \tablecolumns{3}
 \tablewidth{0pt}
 \tablecaption{Proper Motion Catalog\label{tab:catal}}
 \tablehead{
    \colhead{Unit}  &
    \colhead{Label} &
    \colhead{Explanations}
    }
\startdata
 ---    &   ID     & Number, {astrometric catalog ID} \\
 mas~yr$^{-1}$ &   pmx    & weighted proper motion in X \\
 mas~yr$^{-1}$ &   pmy    & weighted proper motion in Y \\
 mas~yr$^{-1}$ & err\_pmx     & error of the weighted proper motion in X \\
 mas~yr$^{-1}$ & err\_pmy     & error of the weighted proper motion in Y \\
 mag    &   $m\_astrom$  & proxy stellar magnitude\\
 ---    &   n1     & number of first-epoch datapoints \\
 ---    &   n2     & number of second-epoch datapoints \\
 ---    &   n\_del & number of deleted datapoints \\
pix     &   max\_res & largest residual in either proper-motion fit\\ 
 pix    &   X      & X-coordinate in ACS/WFC pixels aligned with RA\tablenotemark{a} \\
 pix    &   Y      & Y-coordinate in ACS/WFC pixels aligned with Dec \\
 deg    &   RA  & Right Ascension, decimal degrees (J2000) \\
 deg    &   DEC  & Declination, decimal degrees (J2000) \\
{---}    &   {Pointing}  & {Pointing/field center to which source belongs (1, 3, 5)} \\
\enddata
\tablenotetext{a}{Direction of X-coordinate is opposite to RA}
\tablecomments{Celestial coordinates are on epoch 2016.0. Table is available in machine-readable format at MAST.}
\end{deluxetable}

\eject




\begin{thebibliography}{}

\bibitem[Abadi et al.(2016)]{tensorflow} Abadi, M., Agarwal, A., Barham, P., et al.\ 2016, arXiv:1603.04467


\bibitem[\protect\citeauthoryear{Anderson \& King}{2006}]{anderson_king_2006} Anderson, J., \& King, I.~R.\ 2006, Instrument Science Report ACS 2006-01, (Baltimore: STScI) 

\bibitem[Bastian et al.(2010)]{bastian_2010} Bastian, N., Covey, K.~R., \& Meyer, M.~R.\ 2010, \araa, 48, 339 

\bibitem[Bedin et al.(2008)]{bedin_2008} Bedin, L.~R., King, I.~R., Anderson, J., et al.\ 2008, \apj, 678, 1279

{\bibitem[Bellini et al.(2014)]{bellini_2014} Bellini, A., Anderson, J., van der Marel, R.~P., et al.\ 2014, \apj, 797, 115}

\bibitem[Belokurov et al.(2018)]{belokurov_2018} Belokurov, V., Erkal, D., Evans, N.~W., et al.\ 2018, \mnras, 478, 611 


\bibitem[Bocquet \& Carter(2016)]{pygtc} Bocquet, S. \& Carter, F.~W.\ 2016, The Journal of Open Source Software, 1, 46 

\bibitem[Brown et al.(2014)]{brown_sfh} Brown, T.~M., Tumlinson, J., Geha, M., et al.\ 2014, \apj, 796, 91 


\bibitem[Chabrier(2003)]{chabrier_2003} Chabrier, G.\ 2003, \pasp, 115, 763 

\bibitem[Chabrier(2005)]{chabrier_2005} Chabrier, G.\ 2005, The Initial Mass Function 50 Years Later, 327, 41 


\bibitem[Dall'Ora et al.(2006)]{dallora_2006} Dall'Ora, M., Clementini, G., Kinemuchi, K., et al.\ 2006, \apjl, 653, L109 



\bibitem[Dotter et al.(2008)]{dartmouth} Dotter, A., Chaboyer, B., Jevremovi{\'c}, D., et al.\ 2008, \apjs, 178, 89

\bibitem[Duch{\^e}ne \& Kraus(2013)]{duchene_2013} Duch{\^e}ne, G. \& Kraus, A.\ 2013, \araa, 51, 269

\bibitem[El-Badry et al.(2017)]{elbadry_2017} El-Badry, K., Weisz, D.~R., \& Quataert, E.\ 2017, \mnras, 468, 319 



\bibitem[Filion et al.(2020)]{paper1} Filion, C., Kozhurina-Platais, V., Avila, R.~J., et al.\ 2020, \apj, 901, 82 

\bibitem[Filion \& Wyse(2021)]{filion_wyse} Filion, C. \& Wyse, R.~F.~G.\ 2021, \apj, 923, 218 


\bibitem[Foreman-Mackey et al.(2019)]{emcee} Foreman-Mackey, D., Farr, W., Sinha, M., et al.\ 2019, The Journal of Open Source Software, 4, 1864


\bibitem[Frebel et al.(2016)]{frebel_2016} Frebel, A., Norris, J.~E., Gilmore, G., et al.\ 2016, \apj, 826, 110 


\bibitem[Gaia Collaboration et al.(2016)]{gaia_2016} Gaia Collaboration, Prusti, T., de Bruijne, J.~H.~J., et al.\ 2016, \aap, 595, A1 


\bibitem[\protect\citeauthoryear{Gaia Collaboration et al.}{2018}]{gaia2}Gaia Collaboration, Lindegren, L., Hern\'{a}ndez, J., et al.\ 2018, \aap, 616, A2

\bibitem[Gaia Collaboration et al.(2021)]{gaiaedr3} Gaia Collaboration, Brown, A.~G.~A., Vallenari, A., et al.\ 2021, \aap, 649, A1 

\bibitem[Geha et al.(2013)]{geha_2013} Geha, M., Brown, T.~M., Tumlinson, J., et al.\ 2013, \apj, 771, 29 

\bibitem[Gennaro et al.(2018a)]{gennaro_2018a} Gennaro, M., Tchernyshyov, K., Brown, T.~M., et al.\ 2018, \apj, 855, 20

\bibitem[Gennaro et al.(2018b)]{gennaro_2018b} Gennaro, M., Geha, M., Tchernyshyov, K., et al.\ 2018, \apj, 863, 38 


\bibitem[Girardi et al(2005)]{trilegal} Girardi L., Groenewegen M. A. T., Hatziminaoglou E. and da Costa L. \ 2005, \aap, 436, 895


\bibitem[Hallakoun \& Maoz(2021)]{hallakoun_2021} Hallakoun, N. \& Maoz, D.\ 2021, \mnras, 507, 398 


\bibitem[Harris et al.(2020)]{numpy} Harris, C. R., Millman, K. J., van der Walt, S. J., et al. \ 2020, Nature, 585, 357 


\bibitem[Hayes et al.(2018)]{hayes_2018} Hayes, C.~R., Majewski, S.~R., Shetrone, M., et al.\ 2018, \apj, 852, 49 


{\bibitem[Helmi et al.(2018)]{helmi_2018} Helmi, A., Babusiaux, C., Koppelman, H.~H., et al.\ 2018, \nat, 563, 85}

\bibitem[Helmi(2020)]{helmi_2020} Helmi, A.\ 2020, \araa, 58, 205 


\bibitem[Hunter(2007)]{matplotlib} Hunter, J. D.\ 2007, Computing in Science \& Engineering, 9, 90 



\bibitem[\protect\citeauthoryear{Jenkins et al.}{2021}]{jenkins_2021}Jenkins, S., Li,T.~S., Pace, A.~B., et al.\ 2021, \apj, 920, 92

\bibitem[Kouwenhoven et al.(2009)]{kouwenhoven_2009} Kouwenhoven, M.~B.~N., Brown, A.~G.~A., Goodwin, S.~P., et al.\ 2009, \aap, 493, 979


\bibitem[\protect\citeauthoryear{Kozhurina-Platais et al.}{2018}]{kozhurina_2018}
Kozhurina-Platais, V., Grogin, N., Sabbi, E.\ 2018, Instrument Science Report ACS/WFC 2018-01, (Baltimore, MD: STScI)

\bibitem[Kroupa(2001)]{kroupa_2001} Kroupa, P.\ 2001, \mnras, 322, 231 

\bibitem[Lai et al.(2011)]{lai_2011} Lai, D.~K., Lee, Y.~S., Bolte, M., et al.\ 2011, \apj, 738, 51 



\bibitem[McKinney(2010)]{pandas1} McKinney, W. 2010, Proceedings of the 9th Python in Science Conference, ed. S. van der Walt \& J. Millman (Austin, Tx), 56 



\bibitem[Morton(2015)]{morton_isochrones} Morton, T.~D.\ 2015, Astrophysics Source Code Library. ascl:1503.010


\bibitem[Norris et al.(2010b)]{norris_2010b} Norris, J.~E., Wyse, R.~F.~G., Gilmore, G., et al.\ 2010, \apj, 723, 1632

\bibitem[Okamoto et al.(2012)]{okamoto_2012} Okamoto, S., Arimoto, N., Yamada, Y., et al.\ 2012, \apj, 744, 96


\bibitem[Pace et al.(2022)]{pace_2022} Pace, A.~B., Erkal, D., \& Li, T.~S.\ 2022, arXiv:2205.05699


\bibitem[Pirzkal et al.(2005)]{pirzkal_2005} Pirzkal, N., Sahu, K.~C., Burgasser, A., et al.\ 2005, \apj, 622, 319

\bibitem[\protect\citeauthoryear{Platais et al.}{2015}]{platais_2015}Platais, I.,
van der Marel, R.~P., Lennon, D.~J., et al.\ 2015, \aj, 150, 89


\bibitem[Platais et al.(2020)]{platais_2020} Platais, I., Robberto, M., Bellini, A., et al.\ 2020, \aj, 159, 272 


\bibitem[Reback et al.(2020)]{pandas2} Reback, J., Mendel, B., McKinney, W., et al. 2020, pandas-dev/pandas: Pandas, latest, Zenodo 

\bibitem[Salaris \& Cassisi(2005)]{salaris_2005} Salaris, M., \& Cassisi, S.\ 2005, Evolution of Stars and Stellar Populations (New York: Wiley)


\bibitem[Salpeter(1955)]{salpeter_1955} Salpeter, E.~E.\ 1955, \apj, 121, 161 



\bibitem[Schlafly \& Finkbeiner(2011)]{schlafly_2011} Schlafly, E.~F., \& Finkbeiner, D.~P.\ 2011, \apj, 737, 103

\bibitem[Siegel(2006)]{siegel_2006} Siegel, M.~H.\ 2006, \apjl, 649, L83

\bibitem[Simon(2019)]{simon_review} Simon, J.~D.\ 2019, \araa, 57, 375 

\bibitem[\protect\citeauthoryear{Sohn et al.}{2013}]{sohn_2013} Sohn, S.~T., Besla, G.,
van der Marel, R.~P., et al.\ 2013, \apj, 768:139

\bibitem[Sollima(2020)]{sollima_2019} Sollima, A.\ 2020, \mnras, 495, 2222 

\bibitem[Stetson(1987)]{stetson_1987} Stetson, P.~B.\ 1987, \pasp, 99, 191

\bibitem[STSCI Development Team(2012)]{drizzlepac} STSCI Development Team\ 2012, DrizzlePac: HST image software, ascl:1212.011


\bibitem[Tyson(1988)]{tyson_1988} Tyson, J.~A.\ 1988, \aj, 96, 1 

\bibitem[Virtanen et al.(2020)]{scipy} Virtanen, P., Gommers, R., Oliphant, T. E., et al. 2020, Nature Methods, 17, 261 


\end{thebibliography}
\end{document}